\documentclass[12pt, letterpaper]{article}
\pdfoutput=1
\usepackage{amsmath,amssymb,amsfonts,cite,color,epsf,epsfig,epstopdf,graphics,graphicx,mathtools,subcaption,physics,verbatim}

\usepackage{amsmath,amssymb,amsfonts,cite,color,epsf,epsfig,epstopdf,graphics,graphicx,mathtools,subcaption,physics,verbatim}

\usepackage[svgnames]{xcolor}

\def\thefootnote{*\arabic{footnote}}
\definecolor{ultramarine}{rgb}{0.07, 0.04, 0.56}
\definecolor{cadmiumgreen}{rgb}{0.0, 0.42, 0.24}
\definecolor{indigo(dye)}{rgb}{0.0, 0.25, 0.42}
\usepackage[linktocpage=true,breaklinks]{hyperref}
\hypersetup{
	colorlinks=true,
	citecolor=ultramarine,
	linkcolor=cadmiumgreen,
	urlcolor=indigo(dye),
}

\numberwithin{equation}{section}

\usepackage[utf8]{inputenc}

\usepackage{array}
\newcolumntype{P}[1]{>{\centering\arraybackslash}p{#1}}

\usepackage{dcolumn}
\newcolumntype{M}[1]{>{\centering\arraybackslash}m{#1}}
\newcolumntype{N}{@{}m{0pt}@{}}

\newcommand{\Qe}{\delta_{\cal E}}

\newcommand{\Mpl}{M_{\rm Pl}}

\newcommand{\E}{\mathcal{E}}

\newcommand{\D}{{\rm d}}

\newcommand{\be}{\begin{equation}}  
\newcommand{\ee}{\end{equation}}

\usepackage{tikz}

\begin{document}

\begin{flushright} 
\end{flushright}
\vspace{0.5cm}

\begin{center}

\def\thefootnote{\fnsymbol{footnote}}

{\Large {\bf Abelian and non-Abelian mimetic black holes}}
\\[1cm]
{Mohammad Ali Gorji$^{1}$, Susmita Jana$^{2}$, Pavel Petrov$^{1}$}
\\[.7cm]

{\small\textit{$^{1}$Cosmology, Gravity, and Astroparticle Physics Group, Center for Theoretical Physics of the Universe, Institute for Basic Science (IBS), Daejeon, 34126, Korea}}\\
\vspace{0.25cm}
{\small\textit{$^{2}$Asia Pacific Center for Theoretical Physics, Postech, Pohang 37673, Korea}}

\end{center}

\vspace{.8cm}

\hrule \vspace{0.3cm}

\begin{abstract} 
We investigate black hole solutions in the mimetic extension of the Einstein-Yang-Mills system, in which the Yang-Mills term is constrained to be constant. In the Abelian $U(1)$ case, we find a static spherically symmetric solution that includes the Schwarzschild and Reissner-Nordstr\"{o}m black holes as special cases. Moreover, we identify a stealth Schwarzschild solution with an electric hair. We show that it is impossible to have magnetic hair in the $U(1)$ gauge case, while, in contrast, the non-Abelian $SU(2)$ stealth solutions can sustain both electric and magnetic hair. Unlike the conventional $SU(2)$ Einstein--Yang--Mills black hole, which requires a unit magnetic parameter to exhibit nontrivial non-Abelian contributions, the stealth mimetic $SU(2)$ solution admits genuinely non-Abelian configurations with arbitrary integer magnetic parameter.

\end{abstract}
\vspace{0.5cm} 

\hrule
\def\thefootnote{\arabic{footnote}}
\setcounter{footnote}{0}

\thispagestyle{empty}


\newpage

\section{Introduction}\label{introduction}
General relativity describes gravity very well on intermediate scales, yet indications at the very small and the very large scales suggest it may be incomplete. Theoretically, we expect a quantum theory of gravity to govern short distances and, observationally, astrophysical and cosmological data indicate the existence of dark matter 
\cite{Liddle:1993fq,Bertone:2004pz,Feng:2010gw} and dark energy \cite{SupernovaSearchTeam:1998fmf} with yet unknown origins. Moreover, the cosmological constant problem remains as a fundamental issue at the interface of theory and observation \cite{Weinberg:1988cp,Peebles:2002gy}. These considerations motivate extensions of gravity that may address these issues.

There are many ways to modify gravity, and in most cases extra degrees of freedom appear. The most studied classes are scalar–tensor and vector–tensor theories: pioneering examples include the Brans–Dicke scalar–tensor theory \cite{Brans:1961sx} and Born–Infeld–type vector–tensor models \cite{Born:1934gh}. Higher-derivative interactions were later incorporated within the Horndeski framework \cite{Horndeski:1974wa,Horndeski:1976gi}. The mimetic dark matter scenario was proposed in the context of the scalar–tensor theories in which the conformal mode of gravity plays the role of dark matter \cite{Chamseddine:2013kea,Golovnev:2013jxa}. The mechanism can be formulated systematically as the singular limit of conformal/disformal transformations \cite{Deruelle:2014zza,Arroja:2015wpa,Domenech:2015tca,BenAchour:2024hbg}. Extensions with higher-derivative terms have been developed \cite{Chamseddine:2014vna,Arroja:2015yvd,Cognola:2016gjy,BenAchour:2016cay,Babichev:2016jzg,Chamseddine:2016uef,Chamseddine:2016ktu,Firouzjahi:2017txv,Langlois:2017hdf,Hirano:2017zox,Zheng:2017qfs,Takahashi:2017pje,Gorji:2017cai,Langlois:2018jdg,HosseiniMansoori:2020mxj,Domenech:2023ryc}, and the same idea has also been implemented for gauge fields, leading to gauge-field mimetic models \cite{Barvinsky:2013mea,Gorji:2018okn,Jirousek:2018ago,Gorji:2019ttx,Hammer:2020dqp,Colleaux:2025vtm}. 

In the original mimetic dark matter scenario \cite{Chamseddine:2013kea}, the scalar field acts as a velocity potential: the unit timelike four-velocity is $u^\mu \equiv \partial^\mu \phi$ with $u^\alpha u_\alpha = -1$. The flow is irrotational and geodesic, satisfying $a^\mu \equiv u^\alpha \nabla_\alpha u^\mu = 0$. Geodesic dust congruences generically develop caustic singularities (shell crossing); because $\phi$ encodes the conformal mode of the metric, such singularities signal a breakdown of the gravitational sector in the mimetic theory \cite{Gorji:2020ten,Gorji:2025ajb}. Caustics also appear in numerical $\Lambda$CDM simulations, since cold dark matter is modeled as dust. In the case of $\Lambda$CDM, however, they are not fundamental singularities: they mark the breakdown of the fluid description. In contrast, vector or gauge-field generalizations of the mimetic scenario \cite{Barvinsky:2013mea,Gorji:2018okn,Jirousek:2018ago,Gorji:2019ttx,Hammer:2020dqp,Colleaux:2025vtm} offer a possible solution, as the additional degrees of freedom allow non-vanishing vorticity and, in some cases, nonzero acceleration, which can help prevent formation of the caustics.

The mimetic theory has been extensively explored, with many cosmological solutions reported \cite{Liu:2017puc,Dutta:2017fjw,Brahma:2018dwx,deHaro:2018sqw,deCesare:2018cts,Ganz:2019vre,Myrzakulov:2015qaa}. Beyond cosmology and dark matter, an important avenue for testing modified gravity theories is the study of compact objects, such as static and rotating black holes. Finding black hole solutions in modified gravity is usually more difficult than cosmological solutions. Nevertheless, in recent years, there were build variety of black hole solutions~\cite{Babichev:2016rlq, Babichev:2016fbg,Babichev:2020qpr, Bakopoulos:2022csr, Babichev:2023psy,Bakopoulos:2023fmv,Bakopoulos:2023sdm} for scalar-tensor gravity. Also this direction has been pursued for scalar-tensor mimetic gravity in Refs.~\cite{Myrzakulov:2015kda,Oikonomou:2016fxb,Chen:2017ify,Li:2018uwg,Nashed:2018qag,Sheykhi:2019gvk,Gorji:2020ten}. In the present work, instead of the scalar-tensor mimetic setup, we consider the gauge field mimetic extension of gravity~\cite{Gorji:2018okn,Gorji:2019ttx}. Specifically, we study the Einstein-Yang-Mills action\footnote{We work in units $\hbar = c = 1$ and adopt the metric signature $(-,+,+,+)$.}
\begin{align}
\label{action0}
S = \int \D^4x \, \sqrt{-\det(g_{\mu\nu})} \left[ 
\frac{\Mpl^2}{2} R  
- 2\lambda \left( \tr{{\bf F}_{\mu\nu} {\bf F}^{\mu\nu}} + \epsilon \mathcal{E}^2 \right) 
\right] \,,
\end{align}
where $\Mpl=1/\sqrt{8\pi{G}}$ is the reduced Planck mass, $R$ is the Ricci scalar, $\epsilon=\pm1$, and the constant ${\cal E}$ with dimension $[{\cal E}]=\mbox{mass}^2$ determines the scale of modification. The gauge field strength is defined as usual
\begin{align}
{\bf F}_{\mu\nu} \equiv \partial_\mu {\bf A}_\nu - \partial_\nu {\bf A}_\mu - i g [{\bf A}_\mu, {\bf A}_\nu ] \,,
\end{align}
where ${\bf A}_\mu$ is the gauge potential and $g$ is the gauge coupling. The auxiliary field $\lambda$ enforces the constraint
\begin{align}\label{mimetic-constraint}
{\bf F}_{\mu\nu} {\bf F}^{\mu\nu} = - \epsilon\, \mathcal{E}^2 \,.
\end{align}

The usual Einstein–Yang–Mills system corresponds to the case in which $\lambda$ is a constant, so that $\mathcal{E}$ represents the cosmological constant. So promoting $\lambda$ to an auxiliary field immediately makes the above theory completely different from the usual Einstein-Yang-Mills scenario. 

The gauge group in the Yang-Mills sector can be any compact, semi-simple group, e.g. $SU(N)$. The corresponding Lie algebra will be
\begin{align}
[{\bf T}_a, {\bf T}_b] = i\, f_{abc} {\bf T}_c \,,
\end{align}
where ${\bf T}_a$ are the generators and $f_{abc}$ are the structure constants. The field strength and gauge potential can be expanded in terms of the basis as
\begin{align}
{\bf F}_{\mu\nu} = {\bf T}_a F^a_{\mu\nu} \,,
\qquad
{\bf A}_{\mu} = {\bf T}_a A^a_{\mu} \,.
\end{align}
Cosmological background solution for the spatially curved FLRW with $SU(2)$ gauge symmetry has been found in Ref. \cite{Gorji:2019ttx}. In this paper we analyze static black hole solutions within this framework. We show that the setup admits nontrivial stealth solutions: when the auxiliary field $\lambda$ vanishes around a background configuration, every vacuum solutions of general relativity remains a solution of the mimetic theory. In that case, the Yang–Mills sector propagates on a background identical to its counterpart in general relativity, without affecting the geometry. We focus on two distinct cases corresponding to $U(1)$ and $SU(2)$ gauge groups. For the $U(1)$ case, we investigate both stealth and non-stealth solutions, while for the $SU(2)$ case, we focus on stealth solutions with nontrivial gauge field hair.

Interestingly, the transition from the Abelian to the non-Abelian gauge group drastically alters the physical behavior of the solutions. For instance, in the Abelian case, stealth solutions with pure nontrivial magnetic hair (no electric hair) are impossible, whereas in the $SU(2)$ case, such solutions are allowed. Furthermore, as we will demonstrate, the mimetic setup yields solutions genuinely distinct from the previously found in the  Einstein–Yang–Mills black hole ~\cite{Bizon:1990sr, Volkov:1998cc, Volkov:2016ehx, Gervalle:2025awa}. First, stealth solutions with nontrivial gauge field hair exist; these solutions have a mimetic nature and can be obtained by setting the mimetic field $\lambda$ to zero at the background level. Second, for the $SU(2)$ colored black hole in the usual Einstein-Yang-Mills system, the \emph{magnetic parameter} must be unity; otherwise, the solution reduces to an \textit{embedded Abelian} one~\cite{Volkov:1998cc}. However, for the stealth mimetic $SU(2)$ solution, the configuration remains genuinely non-Abelian for arbitrary integer values of the magnetic parameter. This novel behavior arises from the presence of the mimetic constraint in the action. Consequently, the colored $SU(2)$ as well as the $U(1)$ mimetic black hole solutions exhibit several distinctive and intriguing features absent in standard Einstein–Yang–Mills/Reissner–Nordstr\"{o}m black hole solutions. We expect these features may lead to modifications of black hole thermodynamics, shadows, quasi-normal modes, as well as the emission of gravitational waves from binaries. We leave these vast areas of exploration for future work.

The rest of the paper is organized as follows. In Sec.~\ref{sec: The Abelian $U(1)$ case}, we consider the Abelian $U(1)$ case and we find non-stealth and stealth black hole solutions. We move to the non-Abelian case in Sec.~\ref{sec: Non-Abelian SU(2) case} and we sumamrize our results in Sec.~\ref{sec: Conclusion and Summary}. Some technical computations are presented in Appendix~\ref{app: Match asymptotic for small tr}.

\section{Abelian $U(1)$ case}
\label{sec: The Abelian $U(1)$ case}

In this section we aim to find static spherically symmetric black hole solutions in mimetic gravity theory \eqref{action0} when the Yang-Mills field has $U(1)$ gauge group. In that case, the generators are $1\times1$ matrices ${\bf T}$ and $[{\bf A}_\mu,{\bf A}_\nu]=0$. Fixing the trace as $\tr({\bf T}^2)=1/2$, the action \eqref{action0} reduces to
\begin{equation}\label{action}
S_{U(1)} = \int \D^4 x \sqrt{-\det(g_{\mu\nu})} \left[
\frac{M_{\text{Pl}}^2}{2} R - \lambda \left( F_{\mu\nu} F^{\mu\nu} + 2 \epsilon \mathcal{E}^2 \right)
\right],
\end{equation}
where the $U(1)$ strength tensor is given by 
\begin{align}
F_{\mu\nu} = \partial_\mu A_\nu - \partial_\nu A_\mu \,.
\end{align} 

Varying the action \eqref{action} with respect to the auxiliary field $\lambda$ gives
\begin{align}\label{mimetic-constraint-U(1)}
F_{\mu\nu} F^{\mu\nu} = - 2 \epsilon {\cal E}^2 \,,
\end{align}
which can be also equivalently obtained by restricting \eqref{mimetic-constraint} to the $U(1)$ gauge group. Variation with respect to the metric and gauge field $A_\mu$ gives the Einstein equations and generalized Maxwell equations respectively
\begin{equation}\label{EoM-abstract}
\Mpl^2 G_{\mu\nu} = 4 \lambda F_{\mu }{}^{\alpha } F_{\nu \alpha } \,, 
\qquad
\nabla_{\alpha}\big( \lambda F^{\alpha\nu} \big) = 0 \,,
\end{equation}
where $G_{\mu\nu}=R_{\mu\nu}-g_{\mu\nu}R/2$ is the Einstein tensor. There was another term in the right hand side of the Einstein equations as $\lambda \left( F_{\mu\nu} F^{\mu\nu} + 2 \epsilon \mathcal{E}^2 \right)$ which vanishes after imposing the mimetic constraint \eqref{mimetic-constraint-U(1)}.

We consider the static spherically symmetric metric
\begin{equation}\label{static-metric}
g_{\mu\nu} = \mbox{diag}\big( - \sigma^2 (r)~f(r), f(r)^{-1},  r^2, r^2 \sin^2\theta \big) \,,
\end{equation}
where $\sigma$ is the lapse function which, without loss of generality, can be choosen to be positive $\sigma > 0$. 

In the gauge field sector, we consider  \emph{dyonic ansatz} for $A_\mu$ and also static one for the auxiliary field
\begin{align}\label{dyonic-Amu}
A_\mu = \big(a_0 (r), 0, 0, q_m \cos {\theta}  \big) \,,
\qquad
\lambda=\lambda(r) \,,
\end{align}
where  $a_0$ is the electric potential and $q_m$ represents the magnetic parameter. Note that, despite of the presence of electric and magnetic contributions in the background configuration \eqref{dyonic-Amu}, one can directly confirm that all spacetime scalar combinations involving the strength field tensor such as $F_{\mu\nu}F^{\mu\nu}$, $\tilde{F}_{\mu\nu}F^{\mu\nu}$, and $R_{\mu\nu\alpha\beta}F^{\mu\nu}F^{\alpha\beta}$ are indeed spherically symmetric ensuring the black hole solutions obtained will possess spherical symmetry:
\begin{align}
&F_{\mu\nu}F^{\mu\nu} 
=  \frac{2q_{m}^2}{r^4} - \frac{2a_0'^2}{\sigma^2} \,,
\qquad
F_{\mu\nu}\tilde{F}^{\mu\nu} 
= - \frac{4 q_{m}{}a_0^{\prime}}{r^2\sigma}\;,
\\ \nonumber
&R_{\mu\nu\rho\alpha}F^{\mu\nu}F^{\rho\alpha} 
= \frac{4 q_{m}^2}{r^6} (1 - f) + \frac{2a_0'^2}{\sigma^3} \bigl(\sigma{f}^{\prime\prime}+3 f^{\prime} \sigma^{\prime} + 2 f \sigma^{\prime\prime}\bigr) \;.  
\end{align}

Using the above result in \eqref{mimetic-constraint-U(1)}, we find
\begin{align}\label{eq:EoM-mimetic}
\frac{ a_0'^2}{\sigma^2} - \frac{q_{m}^2}{r^4} 
= \epsilon \mathcal{E}^2 \;.
\end{align}
The above relation shows that the limit $q_m\to0$ is well-defined while the limit $a'_0\to0$ is not well-defined. Therefore we should always assume $a'_0\neq0$. This means pure magnetic-type solution does not exist in this theory. Moreover, looking at the limit of large $r$, we find $\tfrac{ a_0'^2}{\sigma^2} \approx\epsilon\mathcal{E}^2$ which shows that the negative values of $\epsilon$ are not allowed. Hence, for the rest of the analysis for the $U(1)$ case in this section, we set $\epsilon=+1$. Then, Eq.~\eqref{eq:EoM-mimetic} simplifies into:
\begin{align}\label{eq:phiprime1}
\frac{a_0'}{\sigma} = \pm \frac{q_m}{r^2} \sqrt{1 + x^4} \,,
\end{align}
where we have introduced a new characteristic scale $r_{N}$ and a dimensionless variable $x$ as follows
\begin{align}
r_{N} \equiv \sqrt{\frac{q_{m}}{\mathcal{E}}} \,,
\qquad 
x \equiv \frac{r}{r_N} \,,
\end{align}
in terms of which the mimetic effects can be better tracked. Note that $x$ is only a useful quantity to characterize the mimetic effects and we have to work with both $r$ and $x$.

Plugging the ansatz Eq.~\eqref{dyonic-Amu} in the generalized Maxwell equations in \eqref{EoM-abstract}, we find that all the spatial components vanish identically while the temporal component gives
\begin{align}\label{eq:Eom-A0}
&\left(r^2 \lambda \frac{a_0^{\prime }}{\sigma }\right)^\prime = 0\;.
\end{align}

The pure temporal $t-t$ component of the Einstein equation yields
\begin{align}\label{00}
\Mpl^2\left[ 1-(rf)^\prime \right] = 4 r^2\lambda \frac{a_0'^2}{\sigma^2} \,.
\end{align}
Combining the above equation with the radial $r-r$ component of the Einstein equation, we find
\begin{align}\label{11}
\Mpl^2 r^3 f \sigma \sigma^\prime = 0 \,,
\quad
\Rightarrow
\quad
\sigma = c_2 \,,
\end{align}
where $c_2$ is a dimensionless integration constant. Imposing $\sigma=c_2$ in the angular part, we find
\begin{align}
&\Mpl^2 r^2 \left(r^2 f^{\prime}\right)^\prime 
= 8 q_m^2 \lambda \;.
\end{align}
Indeed all equations are symmetric under the redefinition $q_m \to -q_m$. Therefore, it is sufficient to consider only positive values, $q_m > 0$. For the same reason, we also restrict our analysis to $\mathcal{E} > 0$ without loss of generality. 

In the following two subsections, we consider the cases $\lambda(r)\neq0$ and $\lambda(r)=0$ separately. 

\subsection{Solution with $\lambda(r)\neq0$}
\label{subsec: The nontrivial Abelian solution}

We have found that $\sigma$ is constant, as shown in Eq. \eqref{11}, and therefore we can easily integrate Eq.~\eqref{eq:phiprime1} to find the explicit solution for $a_0$. However, we only deal with $a_0'$ in our setup which is given by Eq.~\eqref{eq:phiprime1} after imposing $\sigma$ is constant. Plugging \eqref{eq:phiprime1} in Eq.~\eqref{eq:Eom-A0} we obtain
\begin{equation}\label{eq:lambda}
 r^2 \lambda \frac{a_0^{\prime }}{c_2}  = c_1 \,,
\quad
\Rightarrow
\quad
\lambda = \frac{1}{q_m} \frac{c_1}{ \sqrt{1 + x^4}} \,,
\end{equation}
where $c_1$ is a dimensionless integration constant. Using \eqref{eq:phiprime1} and above results in \eqref{00} and then integrating, we find
\begin{align}\label{fsol2}
f(r) = 1+ \frac{c_3}{r  } +\frac{Q_m}{r^2} \, _2F_1\left(-\frac{1}{2},-\frac{1}{4};\frac{3}{4};-x^4\right) \,;
\qquad
Q_m \equiv \frac{4 c_1 q_m}{\Mpl^2} \,,
\end{align}
where $c_3$ is a dimensionful integration constant, $\;{}_2 F_1(a,b;c;x)$ is the Hypergeometric function, and we have defined $Q_m$ as the dimensionful magnetic parameter.

We now analyze the solutions in three distinct regimes: (a) Far away from the horizon of the black hole: $x \gg 1$, (b) Inside the horizon of the black hole:  $x \ll 1$, and (c) Near the black hole horizon: $x \sim 1$. 

Far away from the black hole, i.e, at $\;r \to +\infty$, one obtains the following asymptotic behavior of the BH solution:
\begin{align}
&f(r) \to  1 +  \frac{c_3}{r} 
- \frac{4 c_1 \mathcal{E}}{\Mpl^2} \Bigl(1 - \frac{2  \Gamma(3/4)^2}{\pi^{1/2} x}\Bigr) \,,\\
& \lambda \to \frac{c_1 }{ \mathcal{E} r^2 }  \left(1 - \frac{1}{2 x^4} \right) \,, 
\qquad \frac{a_0^{\prime }}{c_2} \to \pm \mathcal{E} \left( 1 + \frac{ 1}{2 x^4}\right)\;.
\end{align}
For ${\cal E}=q_m=0$, we recover Schwarzschild black hole and $c_3$ in $f(r)$ characterizes the mass of the black hole. The corrections are two types. If we set ${\cal E}=0$, all corrections disappear. On the other hand, all $x$-dependent corrections are proportional to $q_m$ since $x=r_N/r$ and $r_N=\sqrt{q_m/{\cal E}}$. These magnetic-type corrections become only important for $x\ll1$ or $r\ll{r}_N$. Assuming that $q_m$ is not too large, which is a reasonable choice, we can fix the $t-t$ component of the metric as $g_{00} \to -1$ such that we recover the Minkowski spacetime (asymptotic flatness condition) at large distances $r\to\infty$. Demanding this boundary condition, we obtain the following relation between $c_1$ and $c_2$:
\begin{align}
c_2 = \pm \frac{1}{\sqrt{1 - \Qe}}\,,
\qquad
\Qe \equiv \frac{4 c_1  \mathcal{E}}{\Mpl^2} \,,
\end{align}
where $\Qe$ is the dimensionless quantity. We see that the existence of an asymptotic region at infinity naturally imposes a bound $\Qe<1$. This result suggests to work with the new radial variable 
\begin{align}\label{tilde-r-def}
\tilde{r} \equiv \frac{r}{\sqrt{1 - \Qe}} \,,
\end{align}
in terms of which the spacetime metric takes the form
\begin{align}
\label{metricsol}
ds^2 &= - h(\tilde{r}) dt^2 + \frac{d\tilde{r}^2}{h(\tilde{r})} + (1 - \Qe ) \tilde{r}^2 d\Omega^2 \,,
\end{align}
where we have defined $h(\tilde{r}) \equiv  \frac{1}{1 - \Qe}\;f\left(\tilde{r} \sqrt{1 - \Qe}\;\right)$. Taking the limit $r\to\infty$ we find $h({\tilde r})\approx1+{\cal O}\left(1/{\tilde r}\right)$ and we can identify the coefficient of ${\cal O}\left(1/{\tilde r}\right)$ with the Schwarzschild radius $r_g \equiv 2M/\Mpl^2$. Doing so, we can get rid of $c_3$ in favor of $r_g$ and $h({\tilde r})$ takes the following form
\begin{align}\label{eq: h for U(1)}
h({\tilde r})& = 1 -  \frac{r_{g}{}}{\tilde{r}} 
+ \frac{\Qe}{1 - \Qe} \left( 1+ \frac{\mathcal{F}(x)}{x^2} \right) \;,
\end{align}
where $x=r/r_N={\tilde r}/{\tilde r}_N$ and we have defined
\begin{align}
\mathcal{F}(x) \equiv {}_2F_1\big(- \tfrac{1}{2}, - \tfrac{1}{4}, \tfrac{3}{4}, - x^4\big) -  \frac{2\Gamma(3/4)^2}{\pi^{1/2}} x \,.
\end{align}
The Schwarzschild limit is now clear: for $\Qe=0$, Eq.~\eqref{tilde-r-def} gives ${\tilde r}=r$ and Eq.~\eqref{eq: h for U(1)} gives $h(r)=1-r_g/r$. The asymptotic behavior of the function ${\cal F}(x)$ is
\begin{align}
{\cal F}(x) =
\begin{cases}
1 -  \frac{2\Gamma(3/4)^2}{\pi^{1/2}} x &x\ll1 \,,
\\
- x^2 + \frac{1}{6 x^2}&x\gg1 \,.
\end{cases}
\end{align}
from which we find
\begin{align}
h\big|_{x\ll1} =   \frac{\Qe}{1 - \Qe}\frac{1}{x^2} \,,  
\qquad
h\big|_{x\gg1} = 1 - \frac{r_{g}{}}{\tilde{r}} +  \frac{\Qe}{6 (1 - \Qe)} \frac{1}{x^4} \,.
\end{align}

In the limit $q_m \to 0$ ($Q_m\to0$), the function $h(r)$ in Eq.~\eqref{eq: h for U(1)} takes the following form
\begin{align}
\label{eq: h qm to zero}
h_{q_m \to 0} = 1 - \frac{r_g}{\tilde{r}}\,.
\end{align}
This shows that in the absence of magnetic charge, the metric is locally Schwarzschild while globally there is a factor $1-\Qe$ in front of the angular part. On the other hand, for the case $\mathcal{E} \to 0$ ($\Qe\to0$) and $c_1 > 0$, one recovers Reissner–Nordstr\"{o}m solution with magnetic charge 
\begin{align}\label{RN-BH}
h_{\mathcal{E} \to 0} =  1 -  \frac{r_g}{ {r}} + \frac{Q_m}{{r}^2}\;.
\end{align}

From the metric solution Eq~\eqref{metricsol} we find another important feature: Since we have fixed $h({\tilde r})$ to match the Minkowski metric at $\tilde{r} \gg 1$, there is a modified solid angle measure (solid angle deficit in the case of $0<\Qe<1$)
\begin{align}
\label{deficit}
\Omega = 4\pi ( 1 - \Qe ),
\end{align}
which is evaluated at constant radius and time. In this sense, the solution corresponds to an asymptotically locally flat solution (or asymptotically conical). It is well known that the solid angle deficit is tightly constrained by observational data~\cite{Clifton:2010cn, Jeong:2010ft, Charnock:2016nzm}. Therefore, one naturally expects the value of $\Qe$ to be sufficiently small. However, as can be seen from Eq.~\eqref{eq: h for U(1)}, the parameter $\Qe$ controls the deviation of our new solution from the standard Reissner–Nordstr\"{o}m metric. To obtain measurable physical predictions, one can consider two possible approaches. The first approach is based on the fact that, unlike cosmic strings, our solution does not necessarily lead to a solid angle deficit. The quantity $\Qe$ can, in principle, take either sign. Thus, if one assumes that the total charge $c_1$ vanishes, hence $\Qe=0$ and one may naively expect a compensation of deviations from $4\pi$ in the solid angle measure. A detailed investigation of this idea is left for future work. The second approach involves considering stealth solutions, which generically appear in the scalar-tensor theories of gravity~\cite{Herdeiro:2014goa, Sotiriou:2015pka}. Such solutions have been extensively studied in Refs.~\cite{Babichev:2013cya, Sotiriou:2014pfa, Babichev:2016rlq, Minamitsuji:2018vuw, BenAchour:2018dap, BenAchour:2019fdf, Motohashi:2019ymr}. In the context of mimetic gravity, and by analogy with Ref.~\cite{Gorji:2020ten}, one can obtain a variety of stealth configurations by setting $\lambda = 0$ at the background level. We will further explore this idea in subsection~\ref{subsec: Abelian stealth solution with the electric hair}.

Besides the presence of angle deficit, another interesting property of the metric solution~\eqref{metricsol} is that, for $q_m\neq0$ (or $r_N\neq0$), the Ricci scalar is non-singular 
\begin{align*}
R = - \frac{4 \lambda F^{\mu \nu } F_{\mu \nu }}{\Mpl^2} =  \frac{2\Qe}{\sqrt{r^4+r_N^4}} \,.
\end{align*}
However, the Ricci-squared and Kretschmann scalars remain singular
\begin{align*}
R_{\alpha\beta\mu\nu}R^{\alpha\beta\mu\nu}\big{|}_{r\to0} &=
14 R_{\mu\nu}R^{\mu\nu}\big{|}_{r\to0} =  \frac{56 Q_m^2}{r^8} \,,
\end{align*}
indicating that the solution is not regular.

In summary, the solution \eqref{metricsol}, with $h({\tilde r})$ given by \eqref{eq: h for U(1)}, has three independent parameters $(r_g, q_m, {\cal E})$ where $r_g$ is the Schwarzschild radius, $q_m$ is the magnetic charge, and ${\cal E}$ is the parameter which determines the scale where the modified gravity (mimetic effects) becomes important. In the limit ${\cal E}=0=q_m$, we find the usual Schwarzschild solution, in the limit ${\cal E}\neq0$ and $q_m=0$, we find a local Schwarzschild solution with an angle deficit given by \eqref{deficit} while for ${\cal E}=0$ and $q_m\neq0$, as expected, we recover the Reissner-Nordstr\"{o}m solution \eqref{RN-BH}. 

\begin{figure}[ht] 
	\begin{subfigure}[b]{0.5\linewidth}
		\centering
		\includegraphics[width=0.9\linewidth]{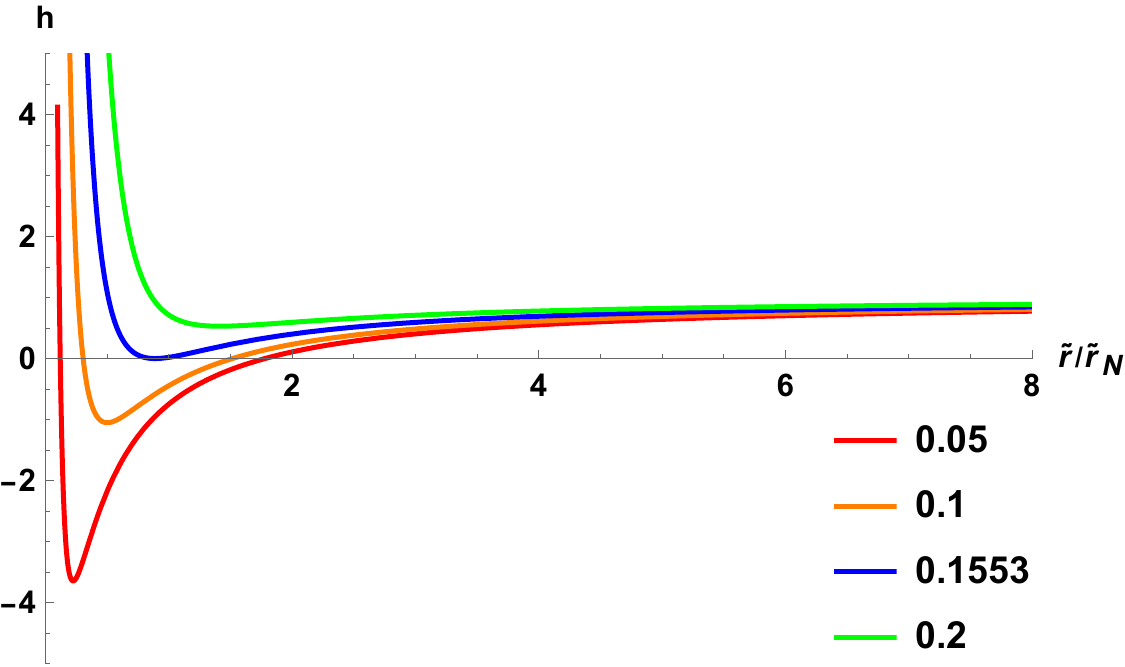}
		\label{fig:function_h+c2}
		\vspace{0ex}
	\end{subfigure}
	\begin{subfigure}[b]{0.5\linewidth}
		\centering
		\includegraphics[width=0.9\linewidth]{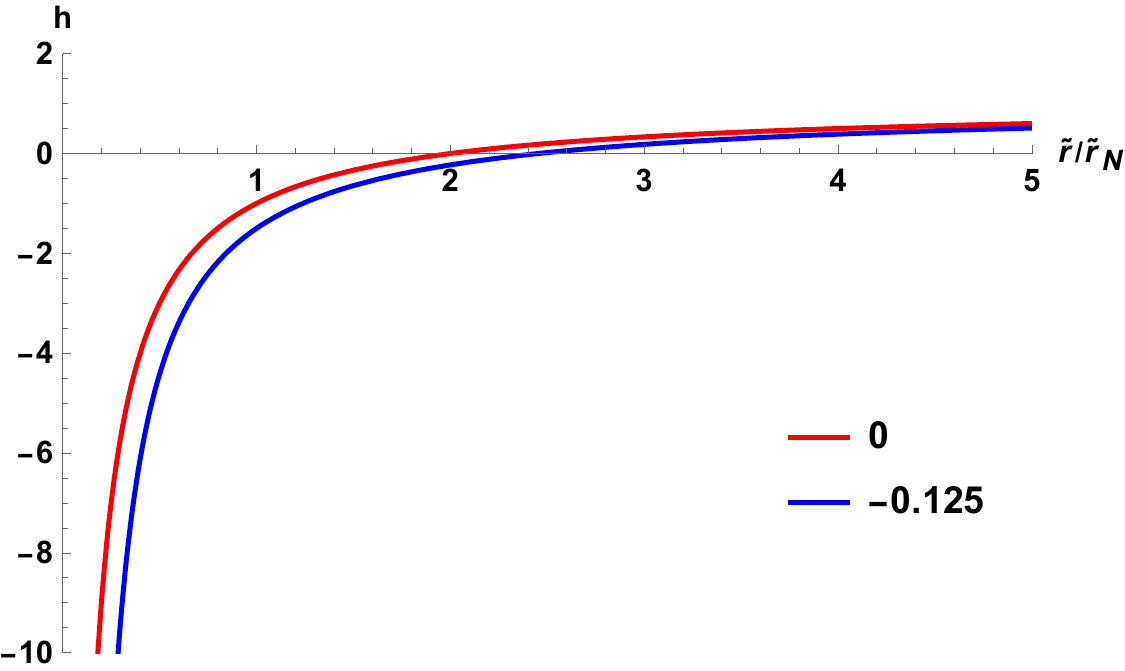}
		\label{fig:function_h-c2}
		\vspace{0ex}
	\end{subfigure} 
	\caption{The function $h$, given by \eqref{eq: h for U(1)}, is plotted for $r_g  = 2,$ $\mathcal{E} = 1$ and $q_m = 1$ (all values here are given in Planck units). Different colors correspond to different values of parameter $c_1$. }
	\label{fig:h}
\end{figure}

The fact that the Reissner-Nordstr\"{o}m balck hole is a subset of our general solution shows that we have at least two horizons in our solution \eqref{eq: h for U(1)}. The question may arise: how many horizon do we have in general? In appendix \ref{app-horizons}, we have proved that for $\Qe<0$, there is only one horizon, whereas for $\Qe>0$, there can be zero (naked singularity), one, or two horizons. These cases are shown in Fig.~\ref{fig:h} where we have plotted the function $h$ for different parameter values of $\Qe$ and $r_g$. We observe that the solution admits cases with two horizons, zero horizons (naked singularity), as well as extreme cases with a single horizon.

\subsection{Stealth solution with $\lambda(r)=0$}\label{subsec: Abelian stealth solution with the electric hair}

Looking at equations \eqref{EoM-abstract}, we find that if $\lambda$ vanishes around a background configuration $\bar{\lambda}=0$, the Einstein equations reduce to that of vacuum while the equation for the $U(1)$ vector field $A_\mu$ trivially satisfies. On the other hand, the equation for the auxiliary field $\lambda$, Eq.~\eqref{mimetic-constraint-U(1)}, implies nontrivial profile for the gauge field $A_\mu$. In particular, for the static spherically symmetric background configuration \eqref{static-metric} and \eqref{dyonic-Amu} we clearly find the Schwarzschild solution 
\begin{align}\label{sigma-f-U(1)}
\sigma(r) = 1 \,,
\quad 
f(r) = 1 - \frac{r_g}{r} \,,
\qquad
\mbox{for}
\quad
\lambda(r) = 0 \,.
\end{align}
Substituting the above solution for $\sigma(r)$ in \eqref{eq:phiprime1} and then integrating, one arrives at the following solution
\begin{align}\label{a0-stealth-U(1)}
a_0(r) = c_3 \mp 
\frac{q_m}{r}\,	{}_2F_1\!\left(-\tfrac{1}{2}, -\tfrac{1}{4}; \tfrac{3}{4}; -x^4\right) \,,
\end{align}
where $c_3$ is an integration constant. For the case of zero magnetic charge, $q_m\to0$, the solution takes a particularly simple form
\begin{align}
a_0(r) = c_3 \pm \mathcal{E} r \;.
\end{align}

The black hole solution \eqref{sigma-f-U(1)} is a stealth solution that coincides identically with the Schwarzschild metric. However, the nonzero electric potential \eqref{a0-stealth-U(1)} indicates that the black hole possesses nontrivial gauge field hair.  This difference will manifest at the level of perturbations, making this stealth solution distinct from the usual Schwarzschild in general relativity. A detailed analysis of perturbations around this black hole background is left for future work.

\section{Non-Abelian $SU(2)$ case }
\label{sec: Non-Abelian SU(2) case}

In the case of $SU(2)$ gauge group, the generators can be chosen as ${\bf T}_a = \frac{1}{2} \tau_a$, and the corresponding structure constants will be $f_{abc} = \varepsilon_{abc}$, with $\tau_a$ denoting the Pauli matrices, and $\varepsilon_{abc}$ being totally antisymmetric tensor with $\varepsilon_{123}=1$. Using $\tr({\bf T}_a,{\bf T}_b)=\delta_{ab}/2$, the action \eqref{action0} can be written in the following form
\begin{equation}\label{action-SU2}
S_{SU(2)} = \int\D^4 x \sqrt{-\det(g_{\mu\nu})} \left[ \,
\frac{\Mpl^2}{2}R  - \lambda \big(  F^{a}_{\mu\nu}F_{a}^{\mu\nu} + 2 \epsilon {\cal E}^2 \big) 
\, \right] \,,
\end{equation}
where 
\begin{equation}
F^{a}_{\mu\nu} = \partial_\mu A^a_\nu - \partial_\nu A^a_\mu + g\, \varepsilon^{a}{}_{bc} A^b_\mu A^c_\nu \,.
\end{equation}
The Greek letters refer to spacetime indices and run as $(\mu = 0,\, 1,\,2,\,3)$ while the Latin index $a$ (where $a = 1, 2, 3$) denotes the group index.

Varying the action \eqref{action-SU2} with respect to the auxiliary field $\lambda$ gives
\begin{align}\label{mimetic-constraint-SU2}
F_{\mu\nu}^a F^{\mu\nu}_a = - 2\epsilon\,\E^2
\end{align}
which can be also equivalently obtained by restricting \eqref{mimetic-constraint} to the $SU(2)$ gauge group.

Variation with respect to metric $g_{\mu\nu}$ and vector field $A_{\mu}^a$ give
\begin{align}\label{eq: SU2 EE and GFE}
& \Mpl^2 G_{\mu\nu} = 4 \lambda F^a_{\mu\alpha}F_{a\nu}{}^{\alpha} \,,
\\ \label{SU2 GFE}
  & \nabla_{\alpha} \left(\lambda F^{a\alpha}{}_{\mu} \right) + \lambda  g \varepsilon^{abc} A^{\alpha }_b F_{c\alpha\mu}  = 0\;.
\end{align}

For the metric, we consider the static spherically symmetric background Eq.~\eqref{static-metric}. The background configuration that is consistent with this configuration is given by the Witten-ansatz \cite{Gervalle:2022npx}
\begin{align}\label{SU2-Ansatz}
\begin{split}
g A^1_\mu &= \left(0,0,\omega_2,-q_m \omega_1 \sin\theta \right) \,,
\\
g A^2_\mu &= \left(0,0,\omega_1,+q_m \omega_2 \sin\theta \right) \,,
\\
g A^3_\mu &= \left(a_0,a_r,\omega_2,+q_m \cos\theta \right) \,,
\end{split}
\end{align}
where $a_0, \, a_r, \, \omega_1, \, \omega_2$ are functions of $r$. The above ansatz is singular in the limit $\theta \rightarrow 0, \pi$. However, this singularity can be removed if we assume that $q_m$ is integer (see Ref. \cite{Gervalle:2022npx} for the details). Therefore, in what follows, we consider $q_m$ to be an integer. Moreover, we choose the following gauge:
\begin{align}
\label{w1w2}
a_r = \alpha^{\prime} \,, 
\qquad
\omega_1 = \omega \cos{\alpha} \,, 
\qquad  \omega_1 = \omega \sin{\alpha} \,,
\end{align}
where $\omega,\, \alpha$ are functions of $r$. One can explicitly confirm that after imposing the above gauge, the gauge-invariant quantities are independent of the choice of $\alpha$. This means that $a_r$ and one of $\omega_{1,2}$ can be removed by fixing the gauge in the static background. Therefore, the two physical variables after fixing the gauge are $a_0$ and $\omega$. By analogy with the $U(1)$ case, we refer to the $a_0$, appearing in Eq.~\eqref{SU2-Ansatz}, as the electric part, and to $\omega$ as the magnetic part. 

Let us first look at the mimetic constraint \eqref{mimetic-constraint-SU2}. Substituting \eqref{static-metric} and \eqref{SU2-Ansatz} in \eqref{mimetic-constraint-SU2} and applying the gauge \eqref{w1w2} we find
\begin{align}\label{eq:EoM-mimetic-SU(2)}
\frac{ a_0'^2}{\sigma^2} 
+ \frac{\left(1+q_m^2\right)}{r^2} \frac{a_0^2}{\sigma^2} \frac{\omega^2}{f}
- \frac{q_{m}^2}{r^4} \left(1-\omega^2\right)^2
- \frac{\left(1+q_m^2\right)}{r^2} f \omega'^2
= \epsilon g^2 \mathcal{E}^2 \;.
\end{align}
For $\omega=0$ the above equation reduces to the $U(1)$ case Eq.~\eqref{eq:EoM-mimetic}. This shows that all non-Abelian effects are encoded in $\omega$ and we, therefore, focus on the case of $\omega\neq0$.


\subsection{Stealth solution with magnetic hair}

Similar to the $U(1)$ case, for $\lambda(r)=0$, the gauge field equations \eqref{SU2 GFE} automatically satisfies. As pointed out in Ref.~\cite{Gervalle:2022npx}, to satisfy the Yang-Mills equation in the usual Einstein-Yang-Mills system, one needs to assume either $q_m^2=1$ or $\omega=0$. Since all non-Abelian effects disappear in the latter case, one has to consider $q_m^2=1$. However, in our case, the fact that $\lambda$ is function of spacetime makes it possible to evade this constraint by assuming $\lambda(r)=0$ while $q_m$ and $\omega$ remain completely arbitrary. This is one of the key findings of this work. Moreover, the energy-momentum tensor in the Einstein equations \eqref{eq: SU2 EE and GFE} vanishes around the static spherically symmetric solution with $\lambda(r)=0$. This means all vacuum solutions of general relativity, like Schwarzschild and Kerr are the solution of our theory as long as $\lambda(r)=0$. For simplicity, we focus on the Schwarzschild case
\begin{align}\label{stealth-metric-SU(2)}
f(r) = 1 - \frac{r_g}{r} \,,
\quad 
\sigma(r) = 1 \,,
\qquad
\mbox{for}
\,\,\,
\lambda(r) = 0 \,.
\end{align}
With the above choices, the only nontrivial equation is the mimetic constraint \eqref{eq:EoM-mimetic-SU(2)} which determines the evolution of the non-Abelian electric and magnetic hairs that are encoded in functions $a_0$ and $\omega$ respectively. For $\omega=0$ all non-Abelian effects disappear and we end up with the result of stealth (Abelian) $U(1)$ Schwarzschild black hole that we already studied in subsection \ref{subsec: Abelian stealth solution with the electric hair}. We thus focus on the case of $a_0=0$ which corresponds to the pure non-Abelian magnetic hair. In that case, Eq. \eqref{eq:EoM-mimetic-SU(2)} simplifies to
\begin{align}\label{eq:EoM-mimetic-SU(2)-magnetic}
- \frac{q_{m}^2}{r^4} \left(1-\omega^2\right)^2
- \frac{\left(1+q_m^2\right)}{r^2} \left(1 - \frac{r_g}{r}\right) \omega'^2
= \epsilon g^2 \mathcal{E}^2 \;,
\end{align}
where we have used Eq. \eqref{stealth-metric-SU(2)}. For $\epsilon>0$, the above equation cannot be satisfied for $r>r_g$ and the exterior region is not well defined. We thus have to consider $\epsilon<0$. It is also important to note that even if we consider $\epsilon<0$, $q_m\neq0$ and $\omega\neq0$ are necessary to have well-defined solution both outside and inside of the stealth black hole solution. Without loss of generality, we choose $\epsilon=-1$, and Eq.~\eqref{eq:EoM-mimetic-SU(2)-magnetic} becomes
\begin{align}\label{eq: for w}
q_{m}^2 \left(1-\omega^2\right)^2
+ \left(1+q_m^2\right)  \left({\tilde r} - 1 \right) {\tilde r} \omega_{{\tilde r}}^2
= \tilde{\cal E}^2 {\tilde r}^4 \,,
\end{align}
where $\omega_{{\tilde r}}$ denotes derivatives of $\omega$ with respect to ${\tilde r}$ and we have defined the dimensionless parameters
\begin{align}
\tilde{r} \equiv \frac{r}{r_g} \,,
\qquad
\tilde{\cal E} \equiv g {\cal E} r_g^2\;.
\end{align}
It is interesting to note that for $g = 0$, hence $\tilde{\cal E}=0$, the situation becomes analogous to the $U(1)$ case, and consequently, no stealth solution with magnetic hair is possible. Hence, we restrict our analysis to nonzero values of $g$.

\begin{figure}[ht] 
	\begin{subfigure}[b]{0.5\linewidth}
		\centering
		\includegraphics[width=0.75\linewidth]{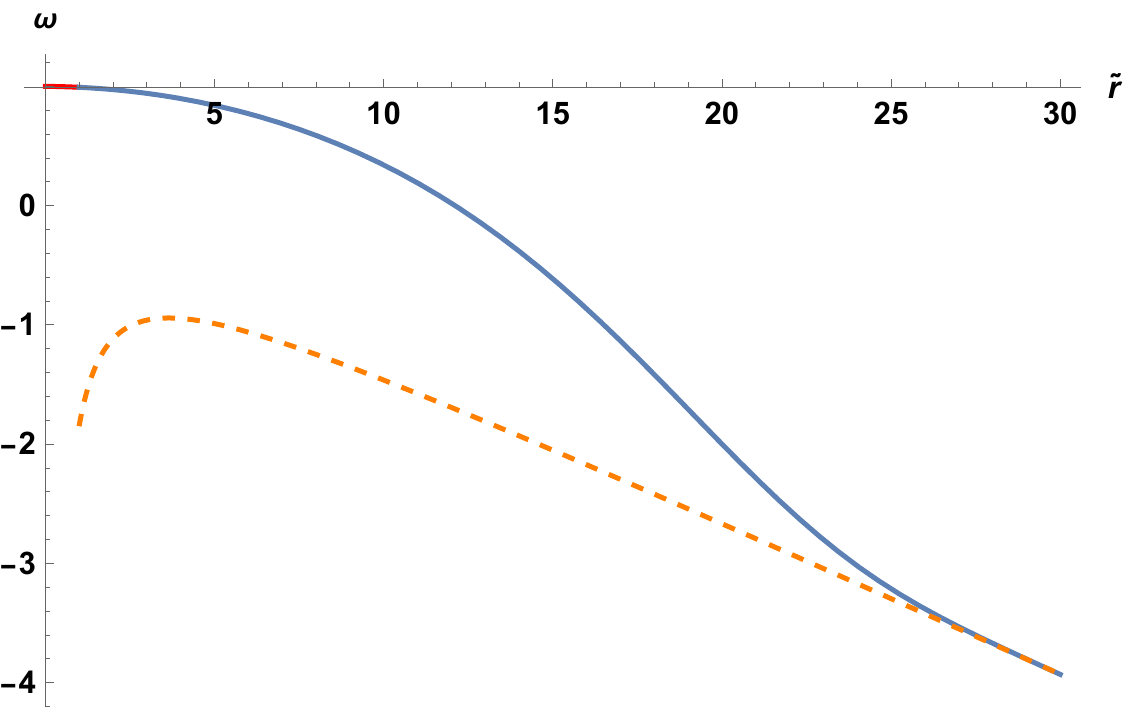}
		\vspace{2ex}
	\end{subfigure}
	\begin{subfigure}[b]{0.5\linewidth}
		\centering
		\includegraphics[width=0.75\linewidth]{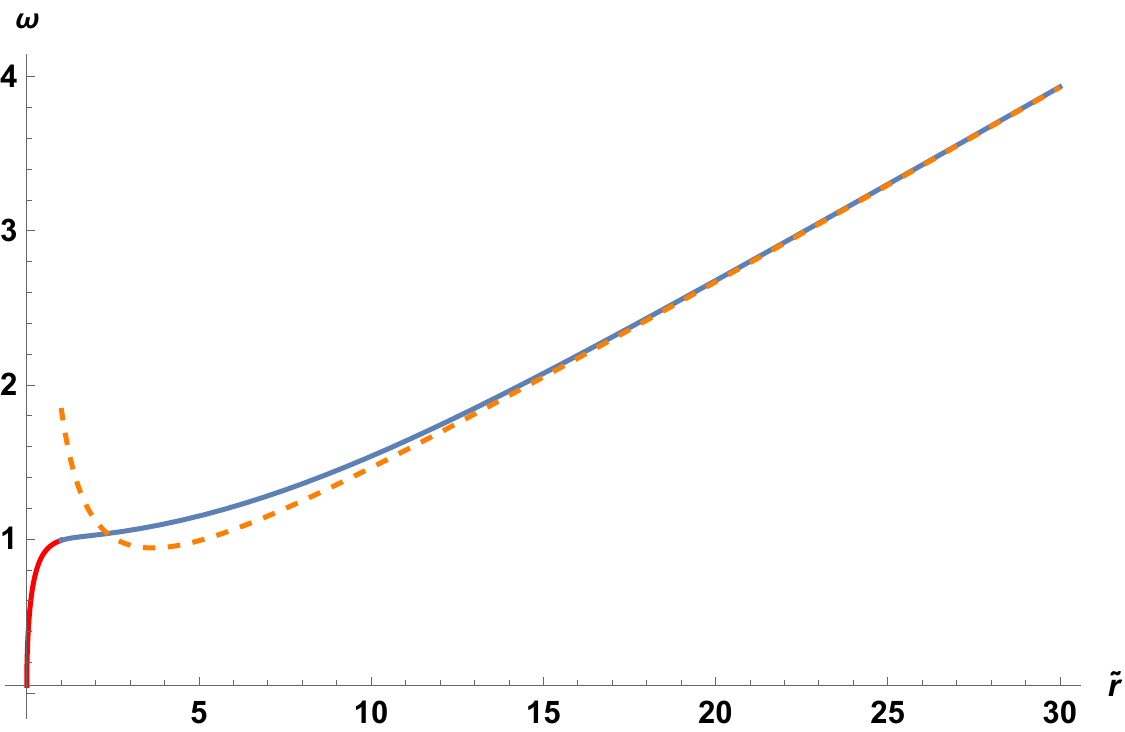} 
		\vspace{2ex}
	\end{subfigure} 
	\begin{subfigure}[b]{0.5\linewidth}
		\centering
		\includegraphics[width=0.75\linewidth]{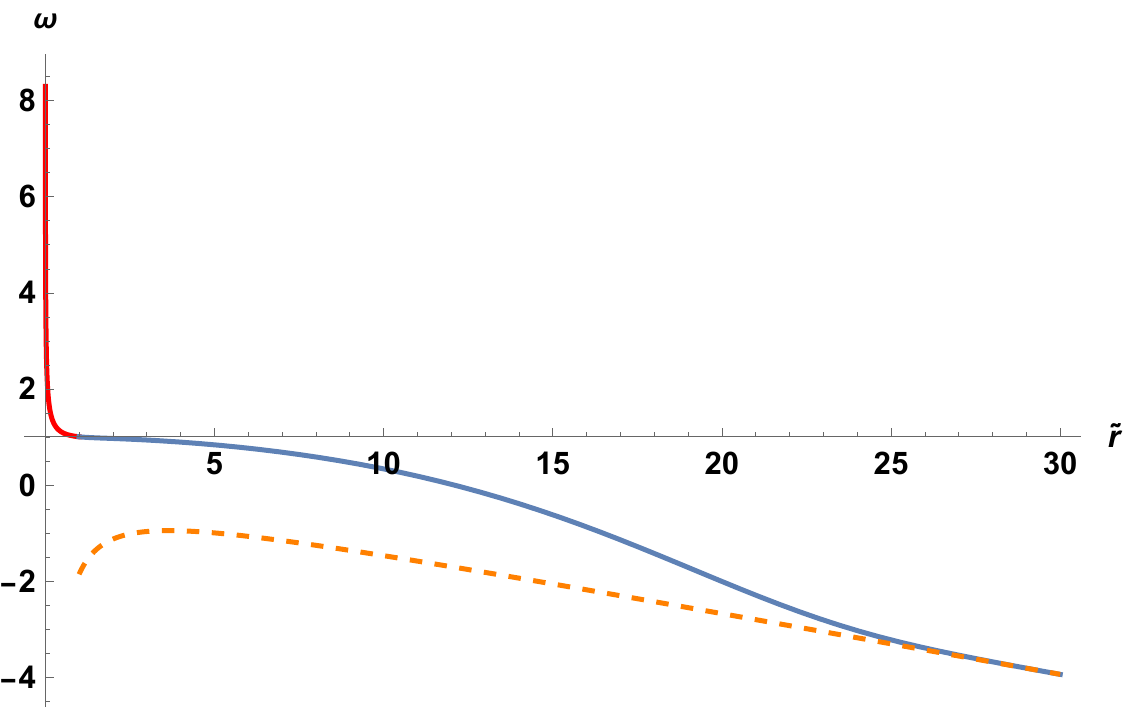} 
	\end{subfigure}
	\begin{subfigure}[b]{0.5\linewidth}
		\centering
		\includegraphics[width=0.75\linewidth]{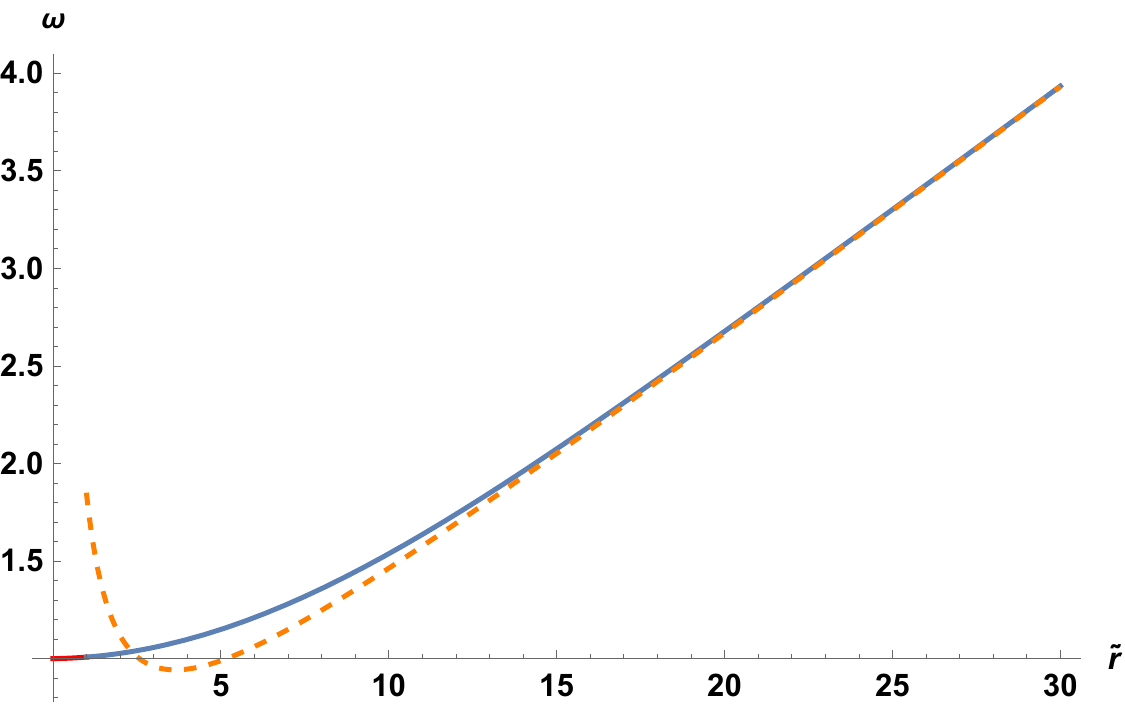} 
	\end{subfigure} 
	\caption{Four distinct numerical solutions for $\omega$ with parameters $q_m = 3$ and $\tilde{\mathcal{E}} = 0.05$. The orange dotted line represents the asymptotic analytical behavior of $\omega$ given by Eq.
		~\eqref{eq: SU2 omega for big tr}.}
	\label{fig:stealth solutions SU2, qm=1}
\end{figure}

Our task is to solve Eq.~\eqref{eq: for w} which does not have an analytical closed form solution. We thus solve it numerically. We see that Eq.~\eqref{eq: for w} is symmetric under the transformation $\omega \to -\omega$ and, therefore, the space of solutions is degenerate. Moreover, the equation possesses a singular point near the horizon at $\tilde{r} = 1$. We set the initial condition for $\omega$ at the horizon. In order to do so, we perform a Taylor expansion to obtain the asymptotic form of the solution in the vicinity of horizon
\begin{align}\label{omega-initial-conditions}
\omega^{\mp}_{\mp} = \omega^{\mp}_0 + \left({\tilde r}-1\right) \omega_{1,\mp}^{\mp} + {\cal O}\left(({\tilde r}-1)^2\right) \,,
\end{align}
where we have defined
\begin{align}
\omega^{\mp}_0 \equiv
\sqrt{1-\tfrac{{\cal E}}{q_m}} \,,
\end{align}
and
\begin{align*}
\omega^{-}_{1,\mp} &=  \frac{2 \tilde{\mathcal{E}} }{1 + q_{m}^2}
\left[q_{m}^{1/2} (q_{m}{} -  \tilde{\mathcal{E}})^{1/2} \mp (1 + 2 q_{m}^2 -  q_{m} \tilde{\mathcal{E}})^{1/2}\right]
\;,
\\
\omega^{+}_{1,\mp} &= \frac{2\tilde{\mathcal{E}}}{1 + q_{m}^2} \left[-q_{m}^{1/2} (q_{m}{} + \tilde{\mathcal{E}})^{1/2} \mp (1 + 2 q_{m}^2 + q_{m}{} \tilde{\mathcal{E}})^{1/2}\right] \;.
\end{align*}
There are four different solutions for $q_m > \tilde{\mathcal{E}}$ while there are two distinct solutions $\omega^{+}_{\mp}$ for $q_m < \tilde{\mathcal{E}}$. On the other hand, for $q_m = \tilde{\mathcal{E}}$, two distinct solutions are $\omega^{+}_{\mp}$ evaluated at $q_m=\tilde{\mathcal{E}}$. However, if we evaluate $\omega^{-}_{\mp}$ at $q_m=\tilde{\mathcal{E}}$, we find $\omega^{-} = \mp \frac{2 \tilde{\mathcal{E}}}{\sqrt{1 + \tilde{\mathcal{E}}^2}} ( \tilde{r} - 1)$ which corresponds to one solution since here we deal with $\omega$ up to a sign. We thus end up with three distinct solutions for the case of $q_m=\tilde{\mathcal{E}}$.

\begin{figure}[ht] 
	\begin{subfigure}[b]{0.5\linewidth}
		\centering
		\includegraphics[width=0.75\linewidth]{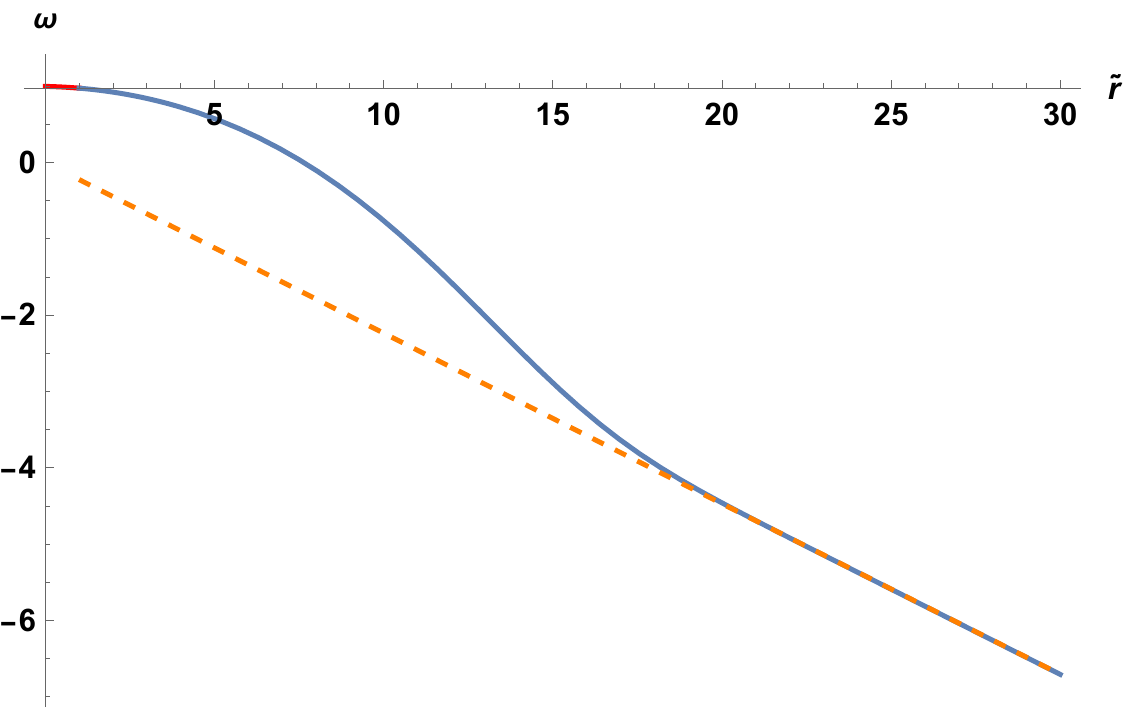}
		\vspace{2ex}
	\end{subfigure}
	\begin{subfigure}[b]{0.5\linewidth}
		\centering
		\includegraphics[width=0.75\linewidth]{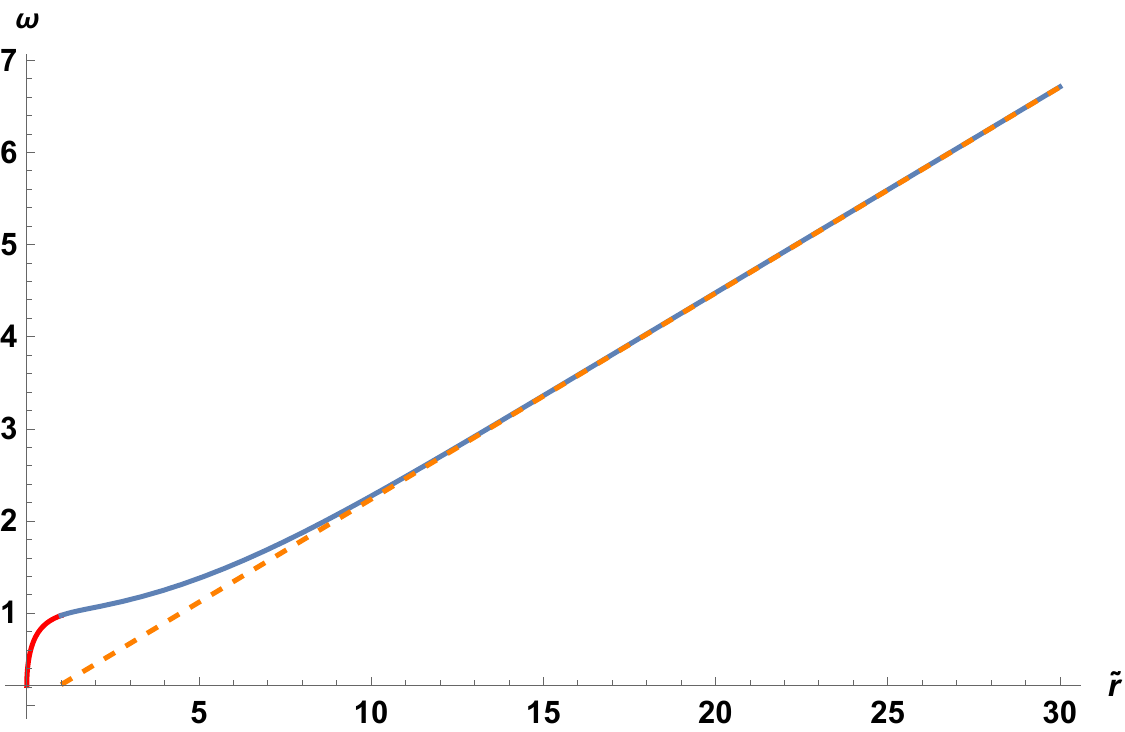}
		\vspace{2ex}
	\end{subfigure} 
	\begin{subfigure}[b]{0.5\linewidth}
		\centering
		\includegraphics[width=0.75\linewidth]{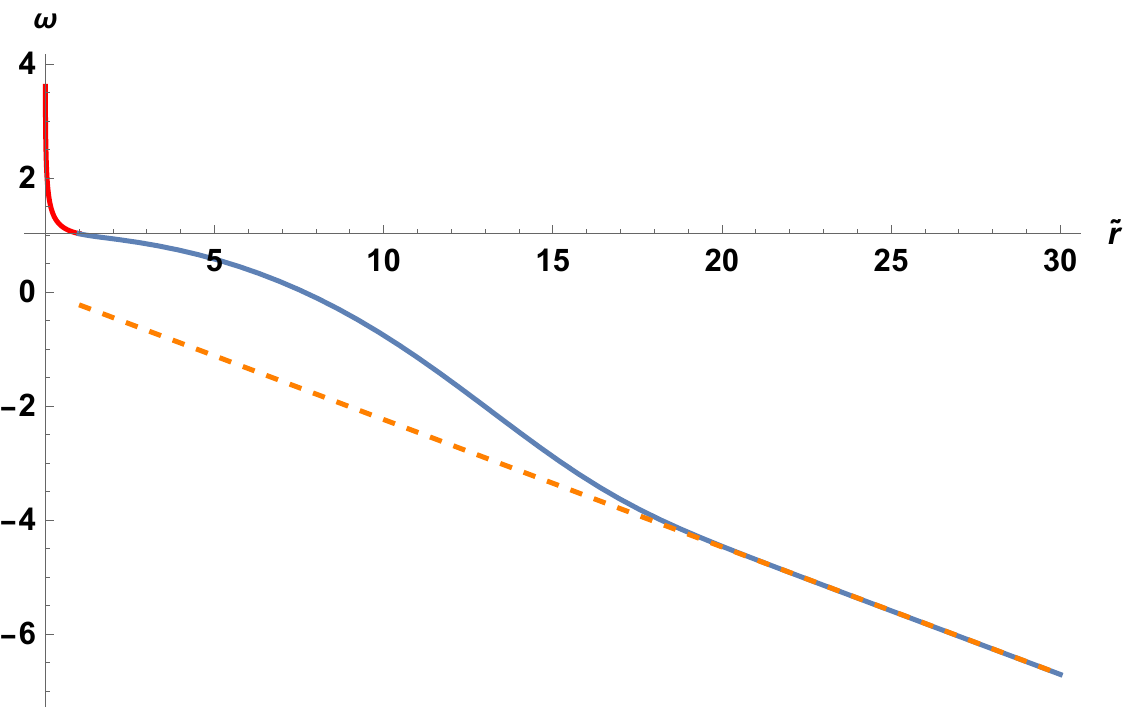} 
	\end{subfigure}
	\begin{subfigure}[b]{0.5\linewidth}
		\centering
		\includegraphics[width=0.75\linewidth]{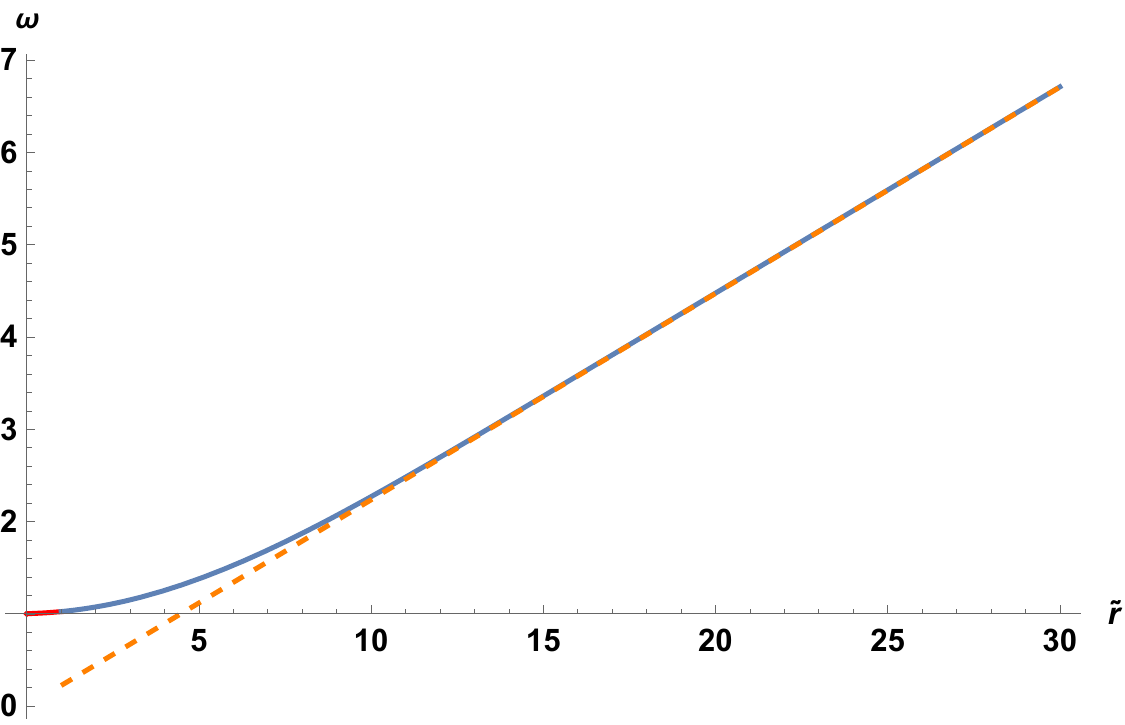} 
	\end{subfigure} 
	\caption{Four distinctive numerical solutions for $\omega$.  The orange dotted lines show asymptotic analytical behavior of $\omega$ given by Eq.~\eqref{eq: SU2 omega for big tr}. For the parameters, we have considered $q_m = 1$ and $\tilde{\mathcal{E}} = 0.05$.}
	\label{fig:stealth solutions SU2, qm=05}
\end{figure}

\begin{figure}[ht] 
	\begin{subfigure}[b]{0.5\linewidth}
		\centering
		\includegraphics[width=0.75\linewidth]{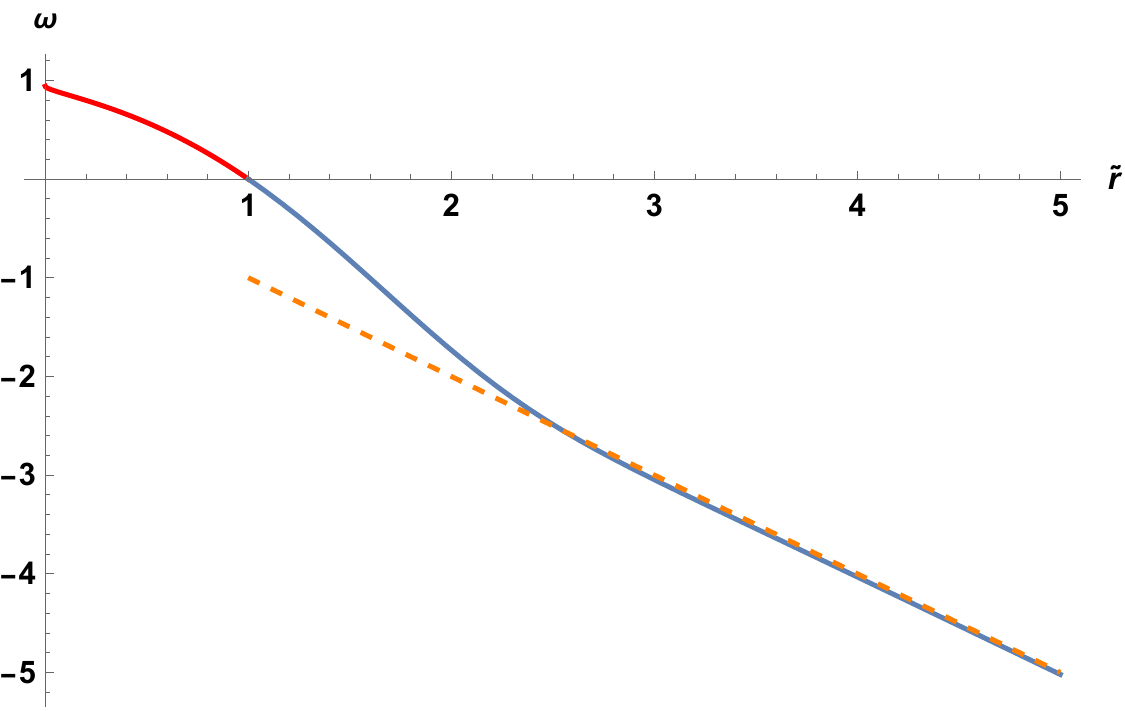}
		\vspace{2ex}
	\end{subfigure}
	\begin{subfigure}[b]{0.5\linewidth}
		\centering
		\includegraphics[width=0.75\linewidth]{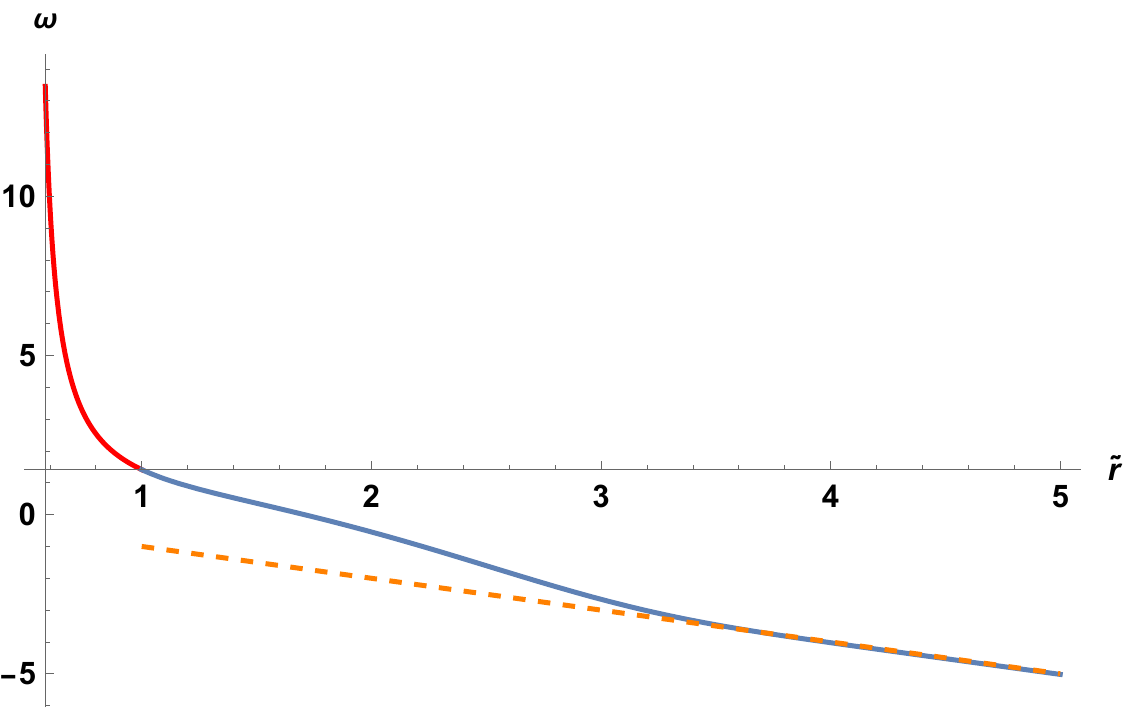}
		\vspace{2ex}
	\end{subfigure} 
	\begin{subfigure}[b]{0.5\linewidth}
		\centering
		\includegraphics[width=0.75\linewidth]{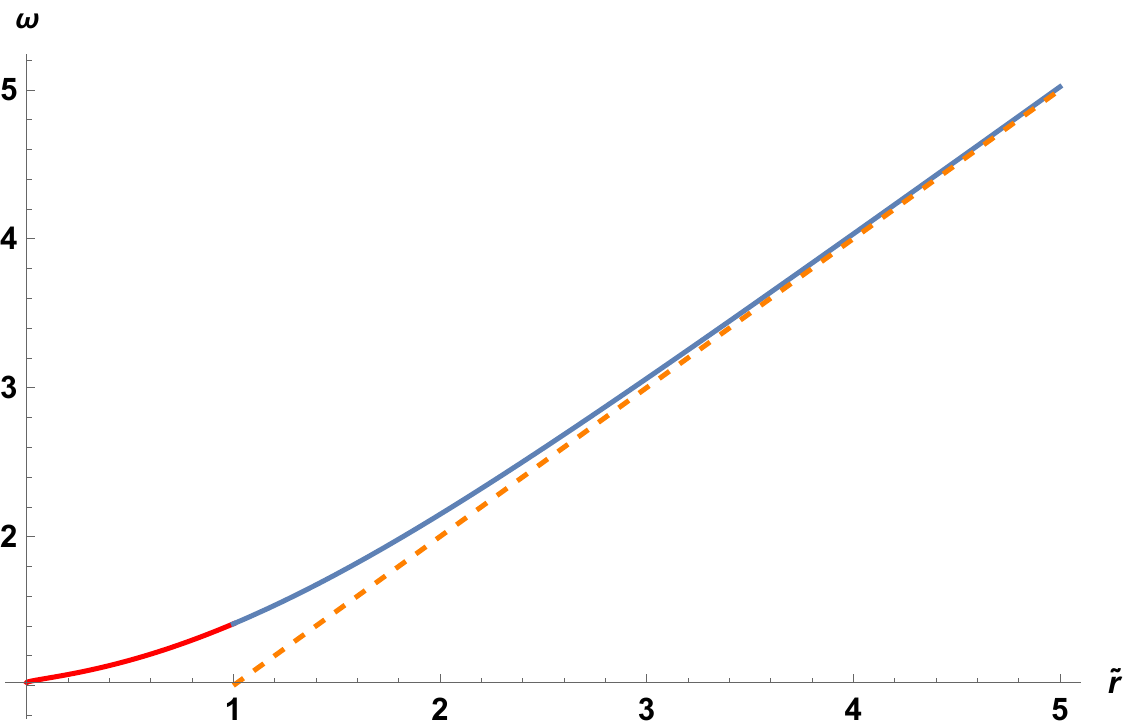} 
	\end{subfigure}
	\caption{Three distinctive numerical solutions for $\omega$. The orange dotted lines show asymptotic analytical behavior of $\omega$ given by Eq.~\eqref{eq: SU2 omega for big tr}. For the parameters, we have considered $q_m  = \tilde{\mathcal{E}} = 1$.}
	\label{fig:stealth solutions SU2, qm=tEE}
\end{figure}

\begin{figure}[ht] 
	\begin{subfigure}[b]{0.5\linewidth}
		\centering
		\includegraphics[width=0.75\linewidth]{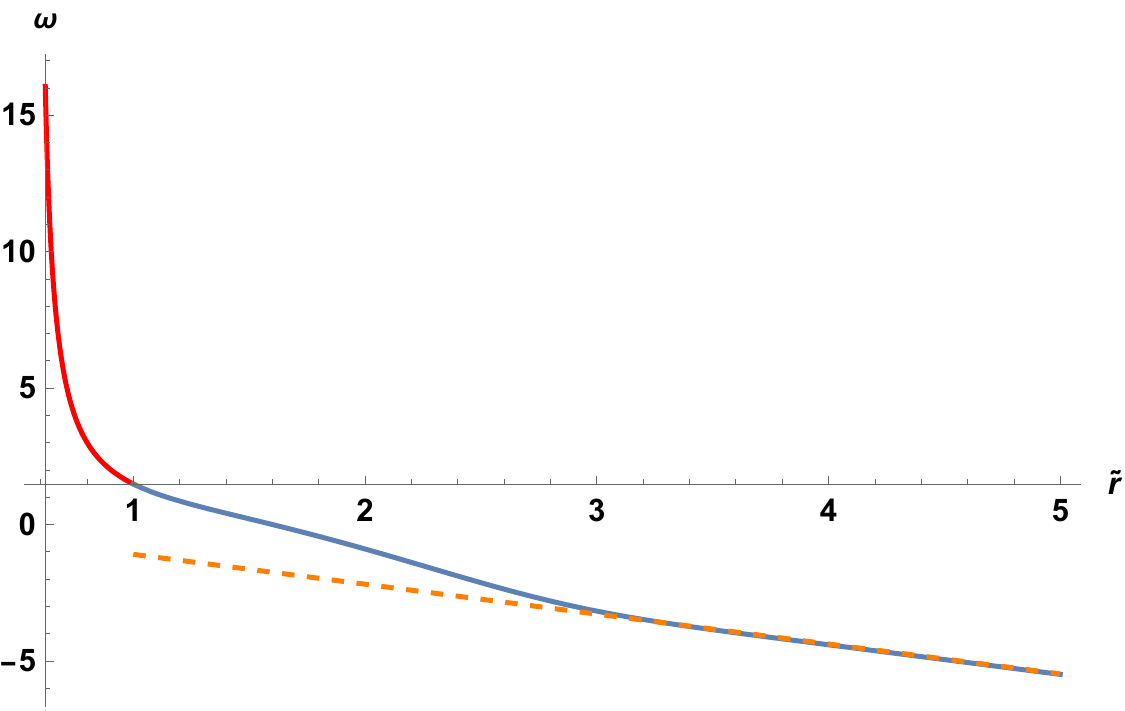}
		\vspace{2ex}
	\end{subfigure}
	\begin{subfigure}[b]{0.5\linewidth}
		\centering
		\includegraphics[width=0.75\linewidth]{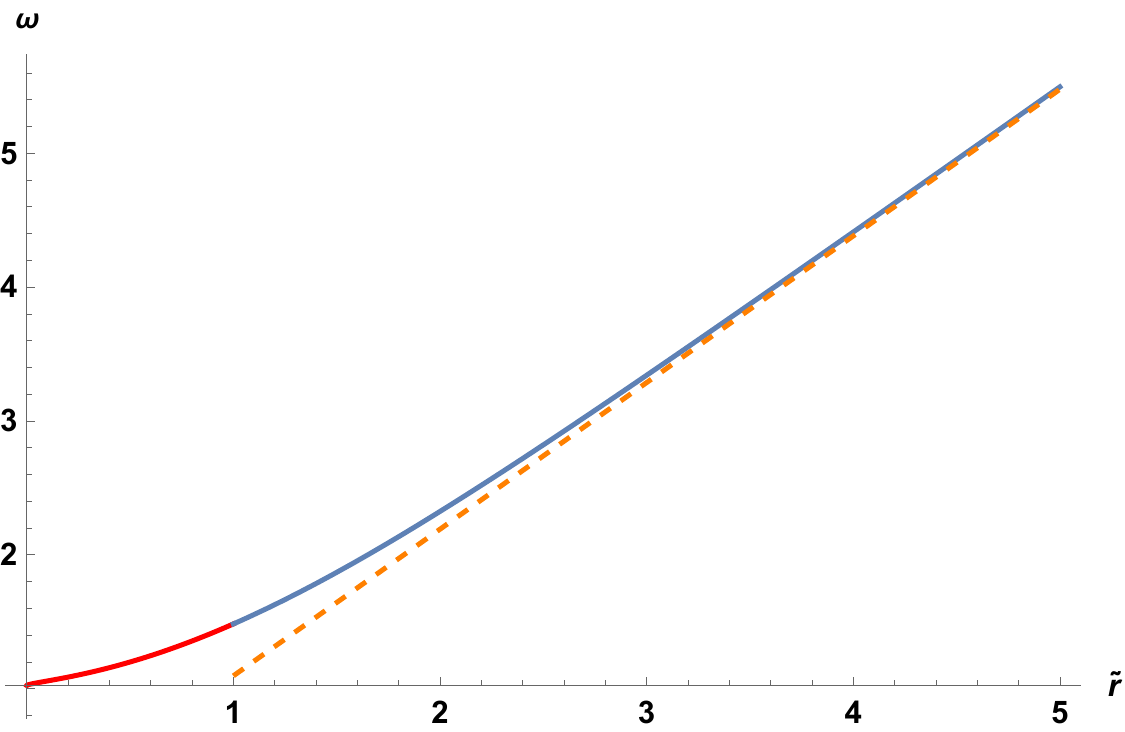}
		\vspace{2ex}
	\end{subfigure} 
	\caption{Two distinct numerical solutions for $\omega$. The orange dotted lines represent the asymptotic analytical behavior of $\omega$ given by Eq.~\eqref{eq: SU2 omega for big tr}.  We have set $q_m = 1$ and $\tilde{\mathcal{E}} = 1.2$.  
	}
	\label{fig:stealth solutions SU2, qm=1/20}
\end{figure}

To further verify our numerical simulation, we derive the analytical asymptotic behavior of $\omega$ for large $\tilde{r}$
\begin{align}
\label{eq: SU2 omega for big tr}
\omega\big|_{\tilde{r} \to \infty} = \pm \sqrt{\tfrac{\tilde{\mathcal{E}}}{q_{m}}} 
\Big( \tilde{r} - \frac{1 -  q_{m}^2}{4 q_{m} \tilde{\mathcal{E}}} \frac{1}{\tilde{r}} \Big)\;.
\end{align}
We explicitly confirm that our numerical solutions converge to the analytical asymptotic behavior in the limits $\tilde{r} \gg 1$ and $\tilde{r} \ll 1$. For completeness, the agreement in the small-$\tilde{r}$ regime is discussed in detail in Appendix~\ref{app: Match asymptotic for small tr}.

The results \eqref{omega-initial-conditions} can be used to set proper initial conditions near the horizon and then solving Eq.~\eqref{eq: for w} numerically. The final results for $\omega$ are plotted in Figs.~\ref{fig:stealth solutions SU2, qm=1}, \ref{fig:stealth solutions SU2, qm=05}, \ref{fig:stealth solutions SU2, qm=tEE}, and \ref{fig:stealth solutions SU2, qm=1/20} where we have confirmed that they all perfectly match the asymptotic expression \eqref{eq: SU2 omega for big tr}. In Figs.~\ref{fig:stealth solutions SU2, qm=1} and~\ref{fig:stealth solutions SU2, qm=05} we have shown the four distinct solutions for the case of $q_m > \tilde{\mathcal{E}}$; the three different solutions for $q_m = \tilde{\mathcal{E}}$ are depicted in Fig.~\ref{fig:stealth solutions SU2, qm=tEE}; and the two solutions for the case of $q_m < \tilde{\mathcal{E}}$ are shown in Fig.~\ref{fig:stealth solutions SU2, qm=1/20}. In the all plots, we set the initial conditions at the points $1 - \delta$ and $1 + \delta$, where $\delta$ is chosen to be sufficiently small; namely, $\delta = 10^{-8}$. We have explicitly verified that the numerical solution converges as the value of $\delta$ is decreased. Therefore, the numerical scheme remains stable and consistent, and in principle, one can choose any sufficiently small value of $\delta$ while ensuring the validity of the numerical simulation.

\section{Summary}
\label{sec: Conclusion and Summary}
In the context of 
mimetic Einstein-Yang-Mills system, we looked for static spherically symmetric black hole solutions. We considered two cases of Abelian $U(1)$ and non-Abelian $SU(2)$ gauge symmetries.

In the $U(1)$ case, which can be interpreted as a mimetic extension of the Einstein-Maxwell theory, considering a dyonic ansatz, we find a new, static spherically symmetric black hole solution given by \eqref{metricsol} and \eqref{eq: h for U(1)}. The solution has three independent parameters $(r_g, q_m, {\cal E})$ where $r_g$ is the Schwarzschild radius, $q_m$ is the magnetic parameter, and ${\cal E}$ is the parameter which determines the scale where the modified gravity (mimetic effects) becomes important. In the limit ${\cal E}=0=q_m$, we find the usual Schwarzschild solution, in the limit ${\cal E}\neq0$ and $q_m=0$, we find a locally Schwarzschild solution with a modified solid angle measure while for ${\cal E}=0$ and $q_m\neq0$, as expected, we recover the Reissner-Nordstr\"{o}m solution. For the case of ${\cal E}\neq0$ and $q_m\neq0$,  we showed that
the solution can admit up to two horizons. For some values of the parameter space, the black hole ceases to exist, and the solution becomes a naked singularity. This behavior is expected, as it has the Reissner-Nordstr\"{o}m black holes as a subset, where sufficiently large electric or magnetic parameter relative to the black hole mass causes a continuous transition from a black hole to an extremal configuration and then to a naked singularity. Although this scenario possesses theoretical interests, it is unlikely to occur in realistic astrophysical settings as the possible electric or magnetic parameter of an astrophysical black hole is expected to be much smaller than its mass. Moreover, we have found another class of configurations, namely stealth solutions, which are black hole solution which have the same metric solution as the general relativity while there is a nontrivial profile of mimetic gauge field. These types of solution can be obtained when the auxiliary mimetic field $\lambda$ vanishes at the background level. 

In the $U(1)$ case, we showed that the stealth solution possesses a nontrivial electric hair, making it physically different from the usual Schwarzschild black hole in general relativity. However, we found that it is impossible to obtain a nontrivial pure magnetic hair for $U(1)$ case. We thus looked for a possible magnetic hair in the case of non-Abelian $SU(2)$ gauge symmetry. Unlike the $U(1)$ case, the $SU(2)$ stealth solution can support both nontrivial electric and magnetic hair. While the metric has the same form as the usual Schwarzschild form, the magnetic hair, encoded in the function $\omega(r)$ which characterizes pure non-Abelian features, has a nontrivial profile. In other words, we have a Schwarzschild black hole with a non-Abelian hair. The non-Abelian hair depends on two parameter: the magnetic parameter $q_m$ and the normalized mimetic scale $\tilde{\cal E} \propto {\cal E}$. We solved the equation for $\omega$ numerically and obtained an analytical expression for its asymptotic behavior, matching the numerical solution in the asymptotic limit. Interestingly, the number of distinct solutions depends on the model parameters: there can exist two (for $q_m < \tilde{\E}$), three (for $q_m = \tilde{\E}$), and four (for $q_m > \tilde{\E}$) different configurations for the gauge field. Whether these branches are stable and physically relevant is not yet known. A detailed stability analysis is left for future work.

Finally, we identify a striking difference between the well-known $SU(2)$ Einstein-Yang-Mills black hole solutions and the stealth $SU(2)$ mimetic configurations. For the $SU(2)$ Einstein-Yang-Mills black hole, the magnetic parameter should be fixed as $q_m^2=1$, otherwise, the solution becomes embedded Abelian ($\omega=0$) and there will be no pure non-Abelian contribution \cite{Volkov:1998cc}. In contrast, for the stealth mimetic $SU(2)$ solution, the solution remains genuinely non-Abelian for arbitrary integer values of the magnetic parameter $q_m$ of the $U(1)$ bundle.

For future directions, it is interesting to study the black hole thermodynamics, shadows, and quasi-normal modes for both $U(1)$ and $SU(2)$ cases. We leave this for future work.

\vspace{0.7cm}

{\bf Acknowledgments:}  We are grateful to  Francesco Di Filippo and Theodoros Nakas for useful discussion and comments. The work of M.A.G. and P.P. is supported by IBS under the project code, IBS-R018-D3. The work of S.J. is supported by the Young Scientist
Training (YST) Fellowship from Asia Pacific Center for Theoretical Physics.
\vspace{0.7cm}

\appendix

\section{Horizons for $U(1)$ black hole}\label{app-horizons}

In this appendix, we show that the black hole solution \eqref{metricsol} with $h({\tilde r})$ given by \eqref{eq: h for U(1)} can have at most two horizons. The Reissner–Nordström black hole is a special case of this solution and already has two horizons, so we know there can be at least two. We then show that no additional horizons can appear.

\begin{figure}[ht] 
	\begin{subfigure}[b]{0.5\linewidth}
		\centering
		\includegraphics[width=0.75\linewidth]{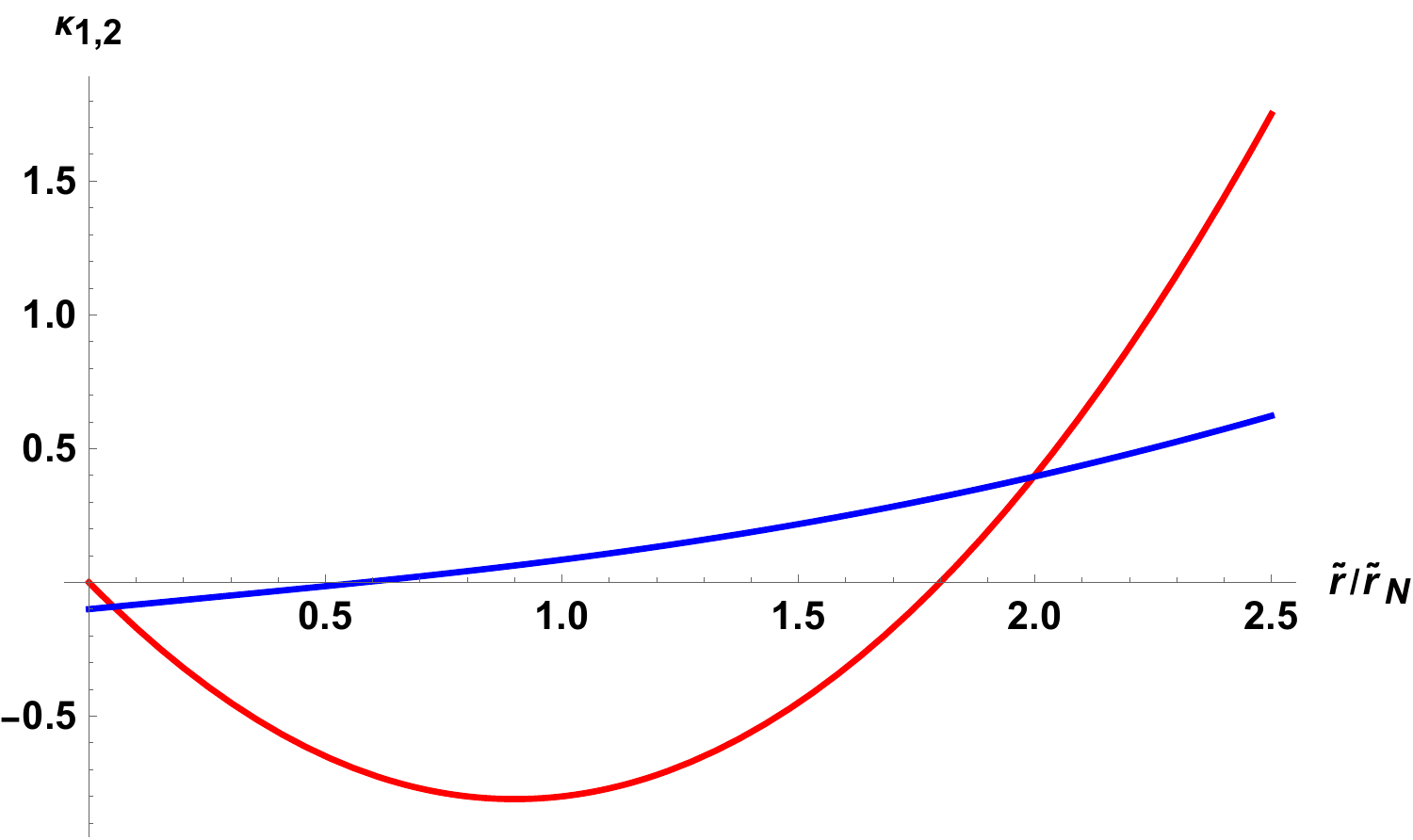}
		\vspace{2ex}
	\end{subfigure}
	\begin{subfigure}[b]{0.5\linewidth}
		\centering
		\includegraphics[width=0.75\linewidth]{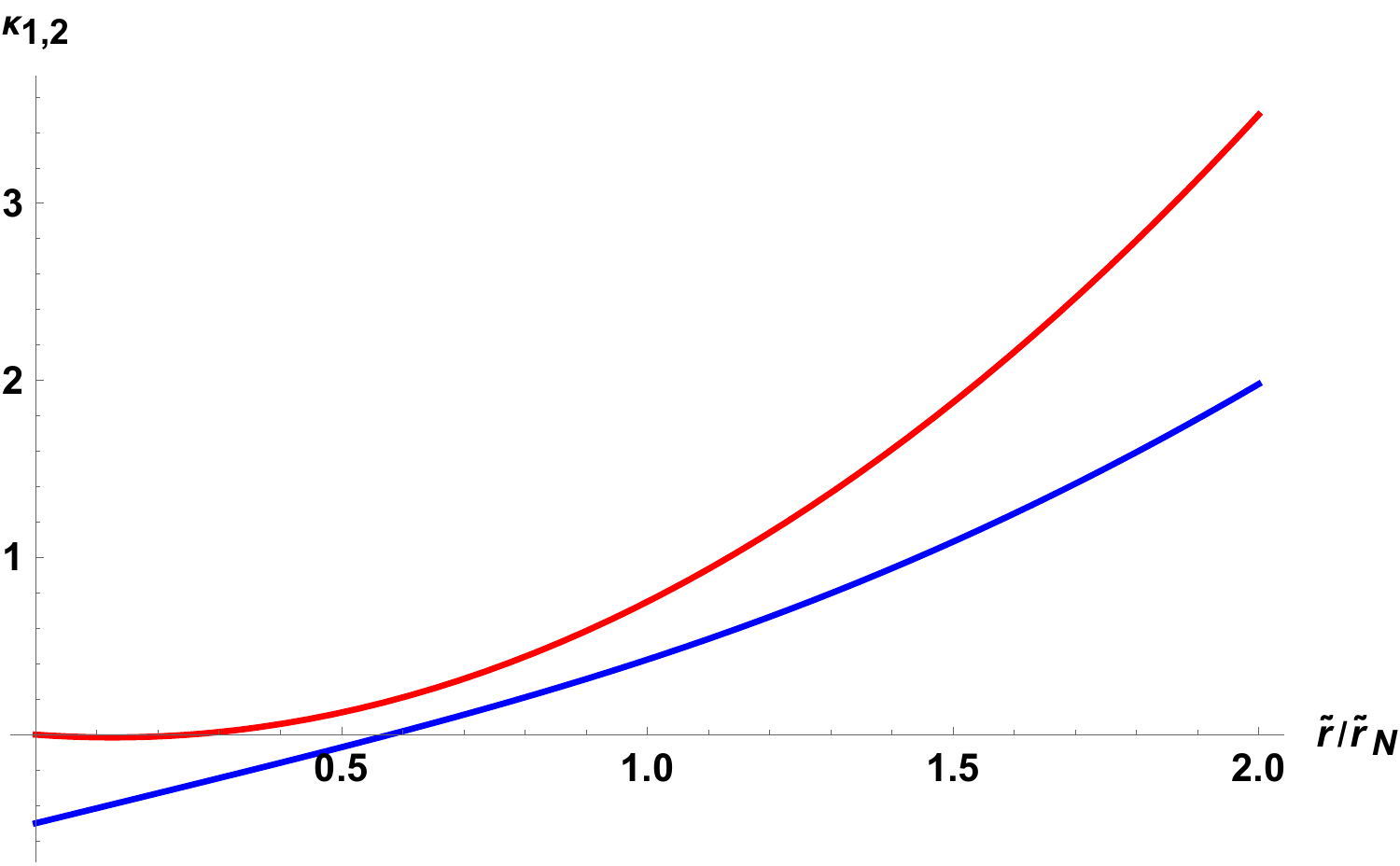}
		\vspace{2ex}
	\end{subfigure} 
	\begin{subfigure}[b]{0.5\linewidth}
		\centering
		\includegraphics[width=0.75\linewidth]{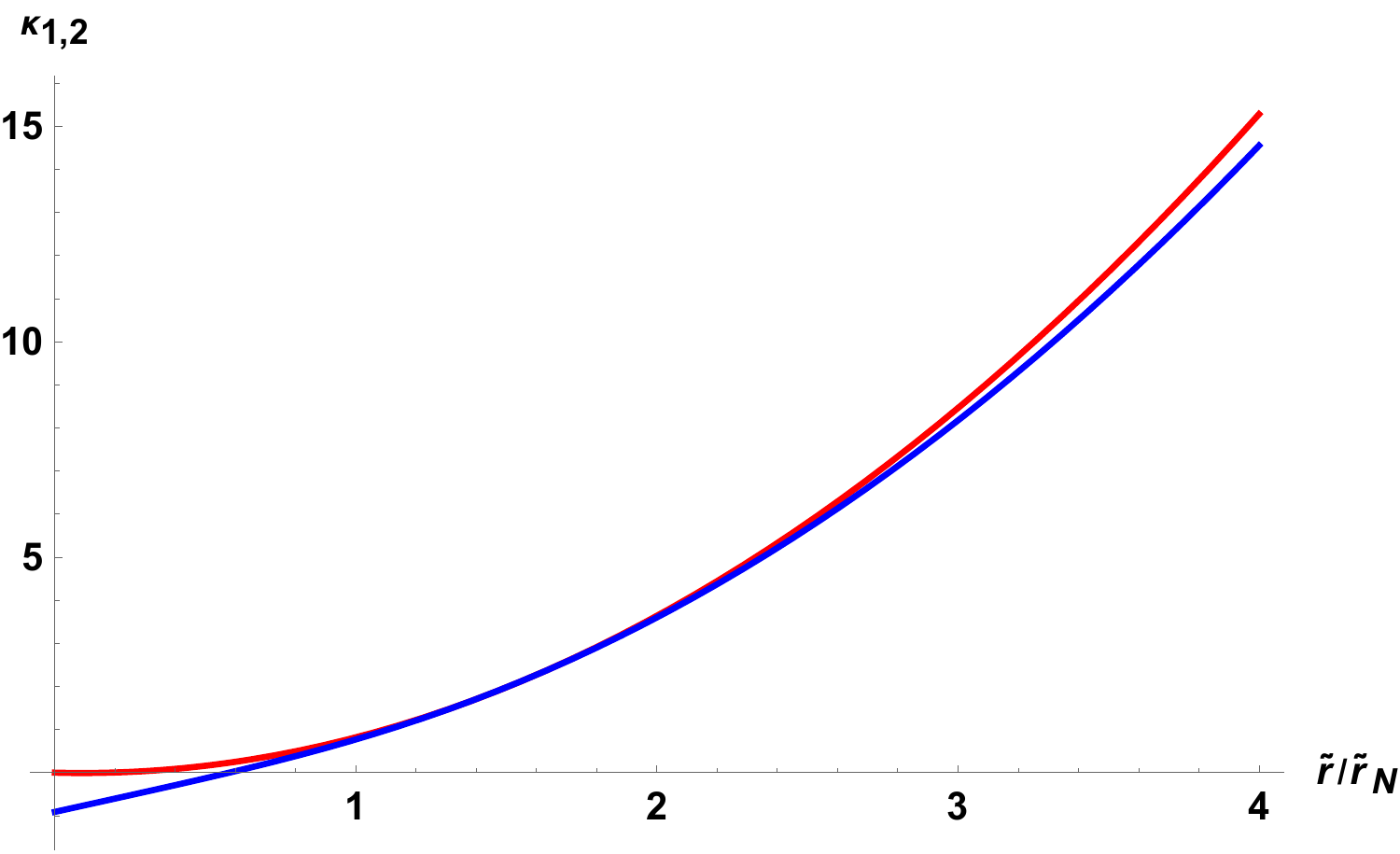} 
	\end{subfigure}
	\begin{subfigure}[b]{0.5\linewidth}
		\centering
		\includegraphics[width=0.75\linewidth]{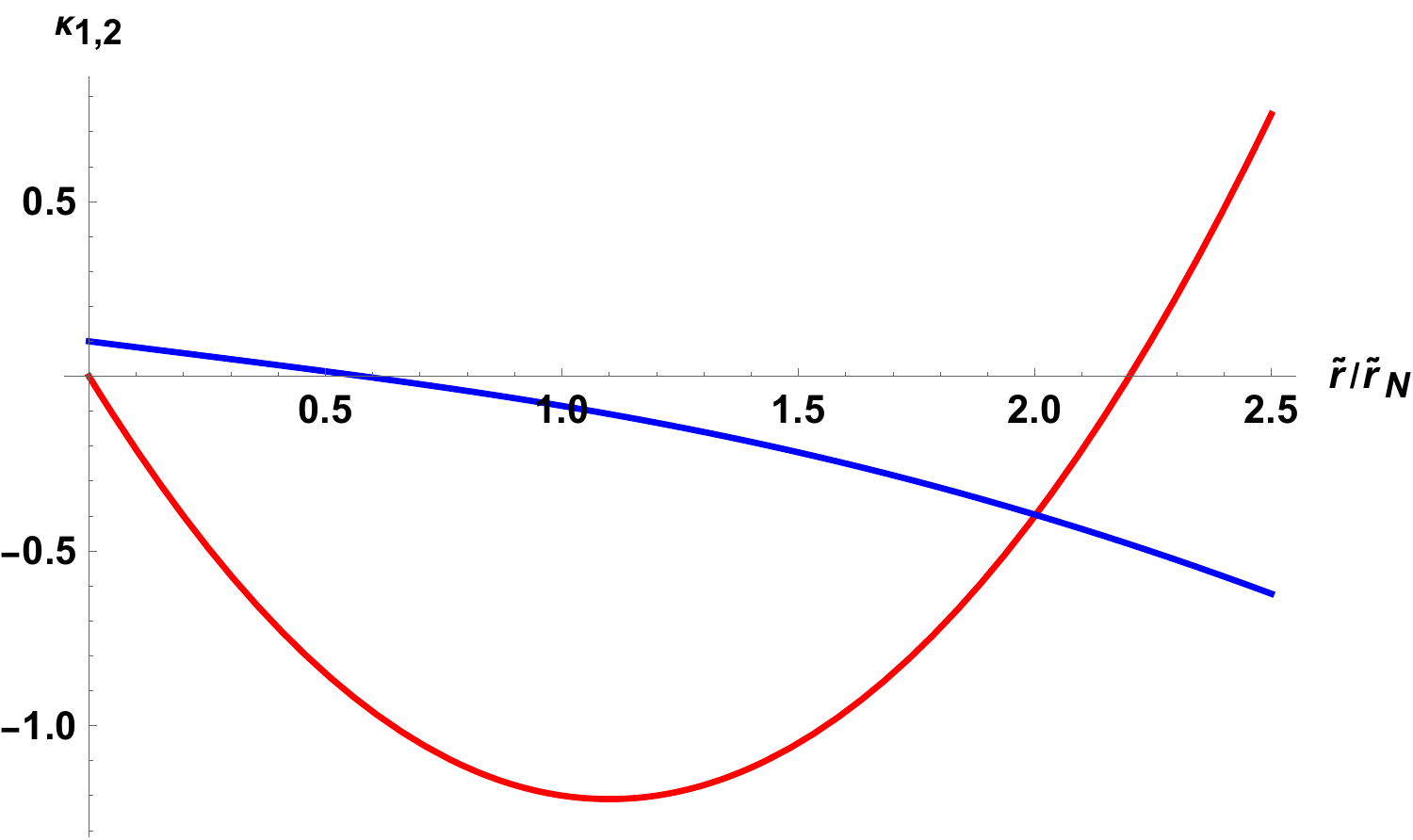} 
	\end{subfigure} 
	\caption{The red lines correspond to function $\kappa_1$ while blue ones correspond to $\kappa_2$.  The up left panel is plotted for $\Qe = 0.1,\;r_g = 2 \tilde{r}_N$ while the up right panel is plotted for $\Qe = 0.5,\;r_g = 0.5 \tilde{r}_N$. The bottom left panel is plotted for the extremal case with one horizon for $\Qe = 0.91,\;r_g = 2 \tilde{r}_N$. Finally, the bottom right panel is plotted for the case of $\Qe < 0$ with $\Qe = -0.1,\;r_g = 2 \tilde{r}_N$. }
\label{fig:k12}
\end{figure}

The condition for the existence of a horizon is $h({\tilde r})=0$. For the practical purposes, we found it more convenient to rewrite it as follows
\begin{align*}
\kappa_1 (x) = \kappa_2 (x) \,,
\end{align*}
where functions $\kappa_{1,2} (x)$ are defined as
\begin{align*}
\kappa_1(x) &= x^2 - (1 - \Qe) \frac{r_{g}}{\tilde{r}_{N}} x \,,
\qquad
\kappa_2(x) = - \Qe\, \mathcal{F}(x)\;.
\end{align*}
The function $\kappa_2$ is monotonic, while the function $\kappa_1$ is parabolic. Therefore, for negative $\Qe$, there is only one horizon, whereas for positive $\Qe$, there can be zero, one, or two horizons. The behavior of the functions $\kappa_1$ and $\kappa_2$ for different parameter values is illustrated in Fig.~\ref{fig:k12}.

\section{Numerical-asymptotic matching for small-$\tilde{r}$ regime} 
\numberwithin{equation}{section}
\label{app: Match asymptotic for small tr}
We have presented the full numerical solution of Eq.~\eqref{eq: for w} for the magnetic hair, that is characterized by $\omega$, where we also showed that the numerical solution agrees well with the asymptotic results for large-$\tilde{r}$. In this appendix, for completeness, we also show that the numerical solution matches the asymptotic results for small-$\tilde{r}$.

\begin{figure}[ht] 
	\begin{subfigure}[b]{0.5\linewidth}
		\centering
		\includegraphics[width=0.75\linewidth]{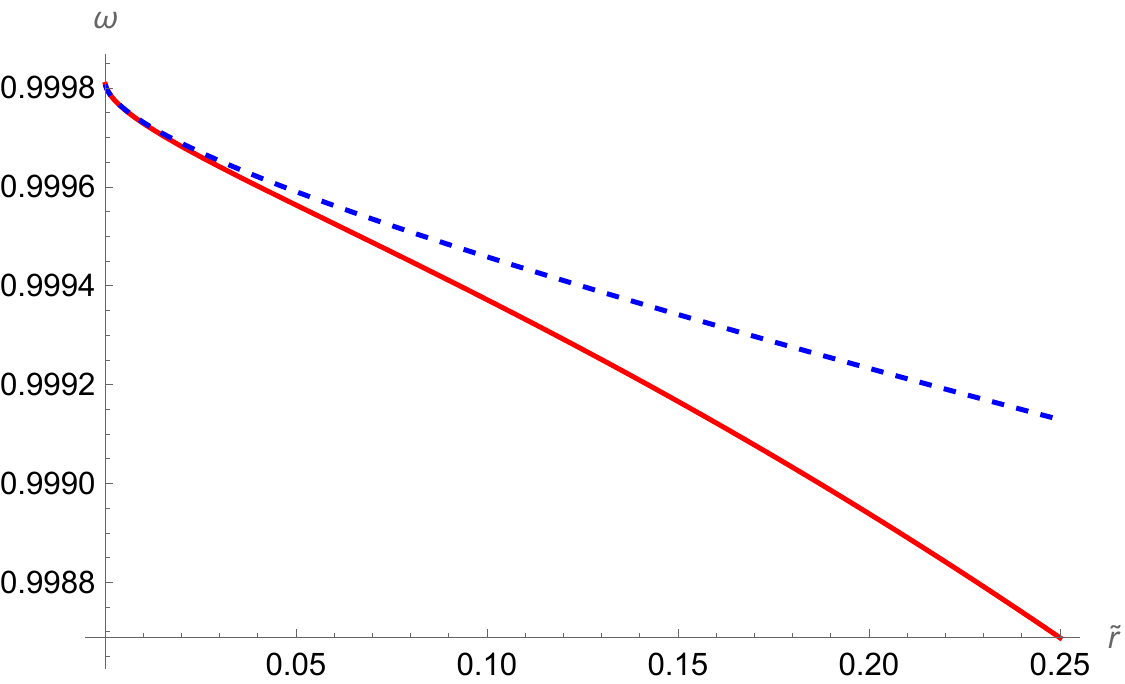}
		\vspace{2ex}
	\end{subfigure}
	\begin{subfigure}[b]{0.5\linewidth}
		\centering
		\includegraphics[width=0.75\linewidth]{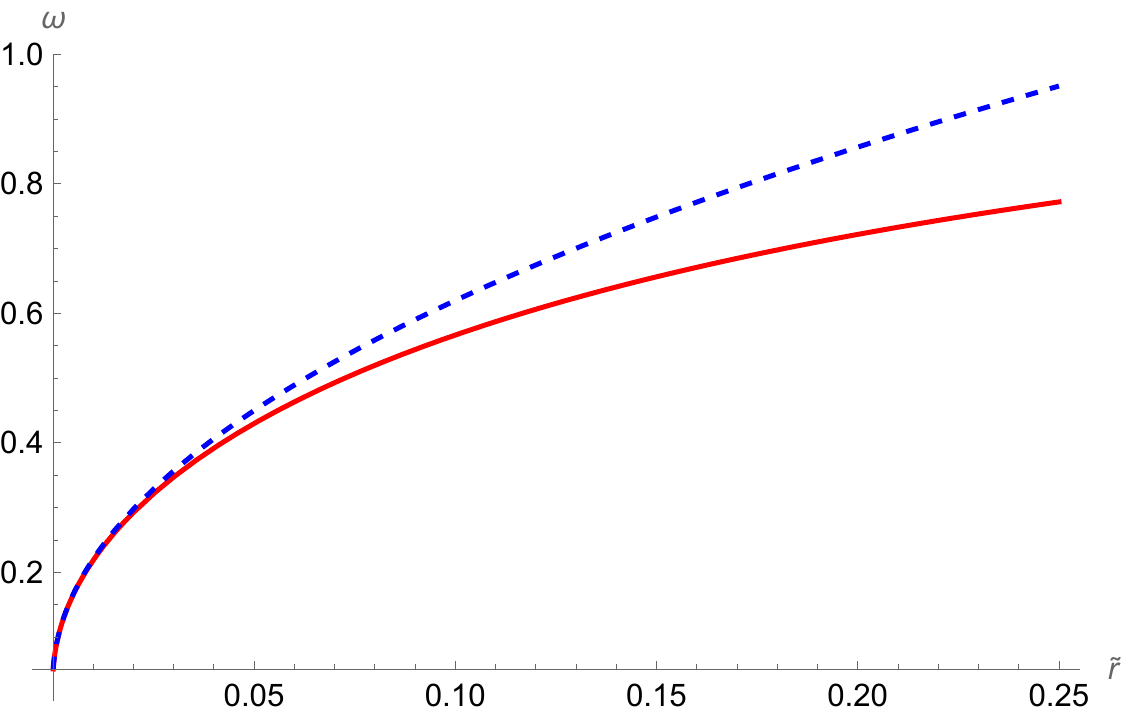}
		\vspace{2ex}
	\end{subfigure} 
	\begin{subfigure}[b]{0.5\linewidth}
		\centering
		\includegraphics[width=0.75\linewidth]{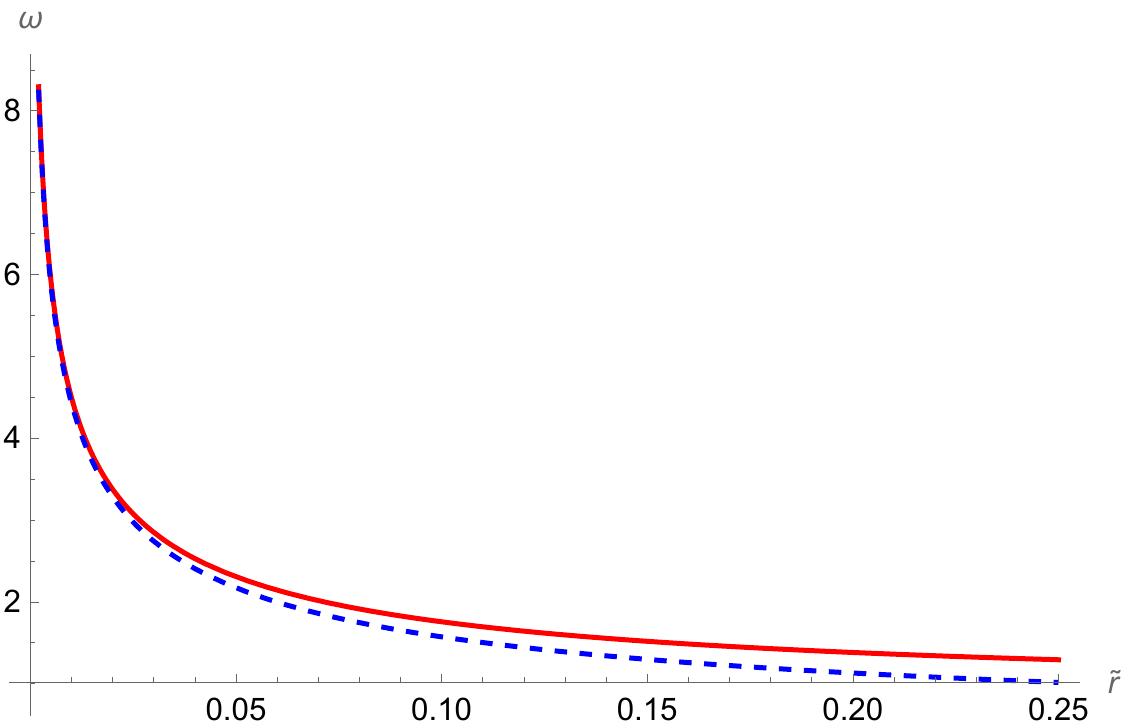} 
	\end{subfigure}
	\begin{subfigure}[b]{0.5\linewidth}
		\centering
		\includegraphics[width=0.75\linewidth]{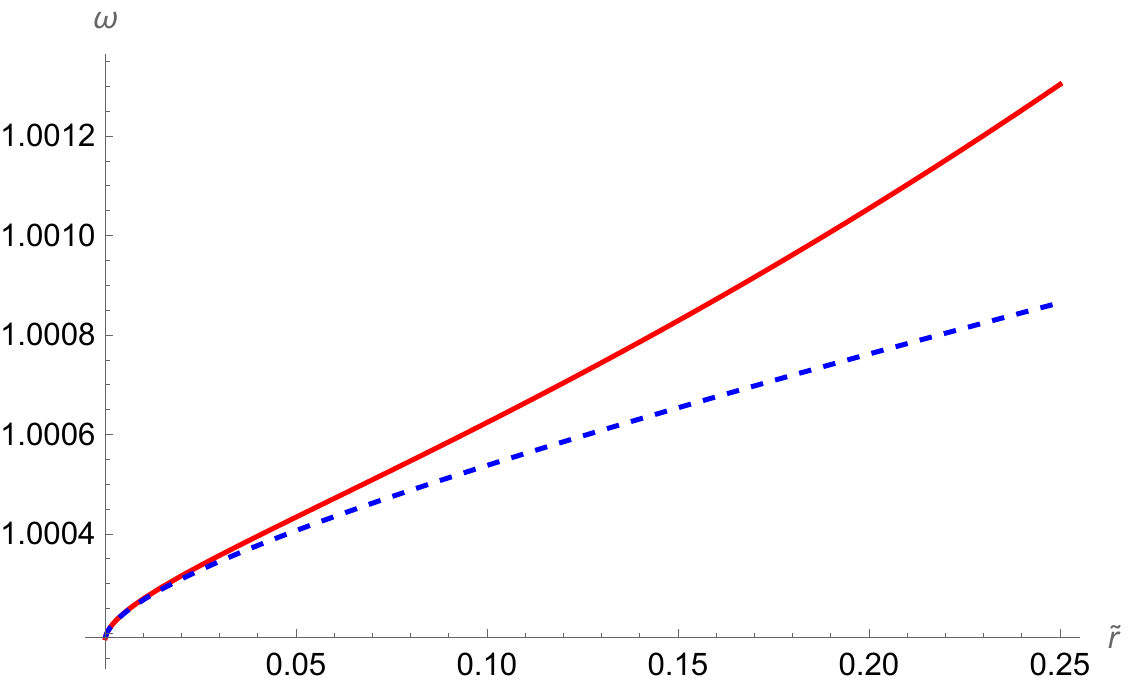} 
	\end{subfigure} 
 \caption{Four distinctive numerical solutions of $\omega$ are plotted for $q_m = 3,\;\tilde{\mathcal{E}} = 0.05$. The blue dotted lines show the asymptotic behavior given by Eq.~\eqref{eq: SU2 omega for small tr}.}
\label{fig: mathc at zero}
\end{figure}

\begin{figure}[!ht]
	\begin{subfigure}[b]{0.5\linewidth}
		\centering
		\includegraphics[width=0.75\linewidth]{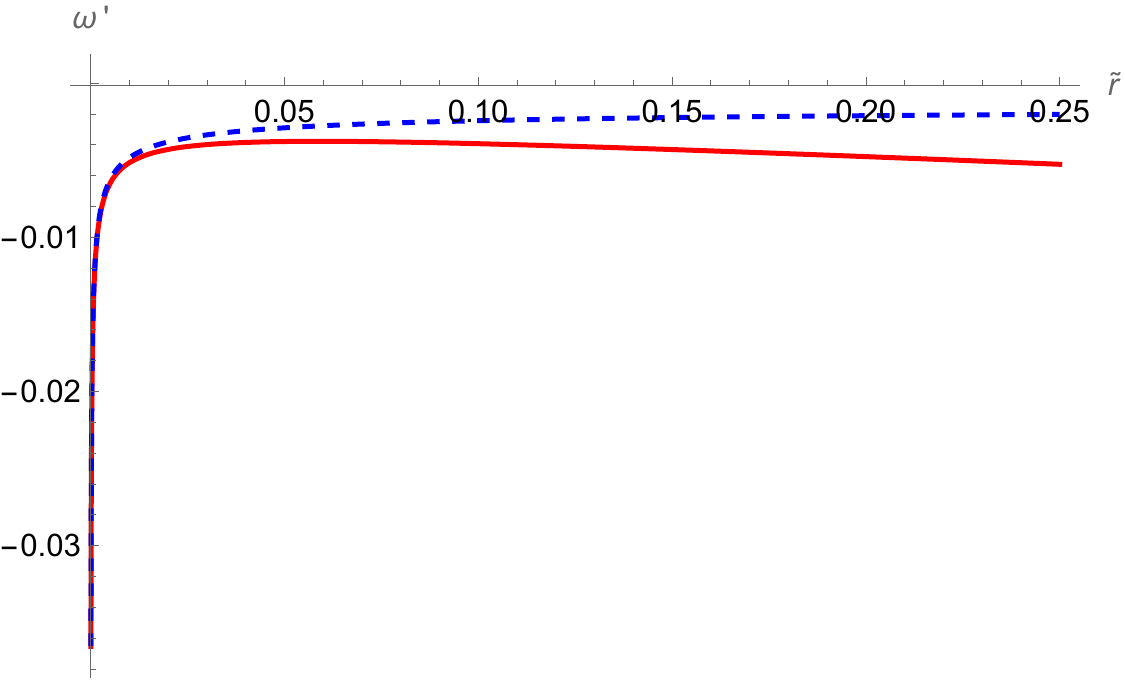}
		\vspace{4ex}
	\end{subfigure}
	\begin{subfigure}[b]{0.5\linewidth}
		\centering
		\includegraphics[width=0.75\linewidth]{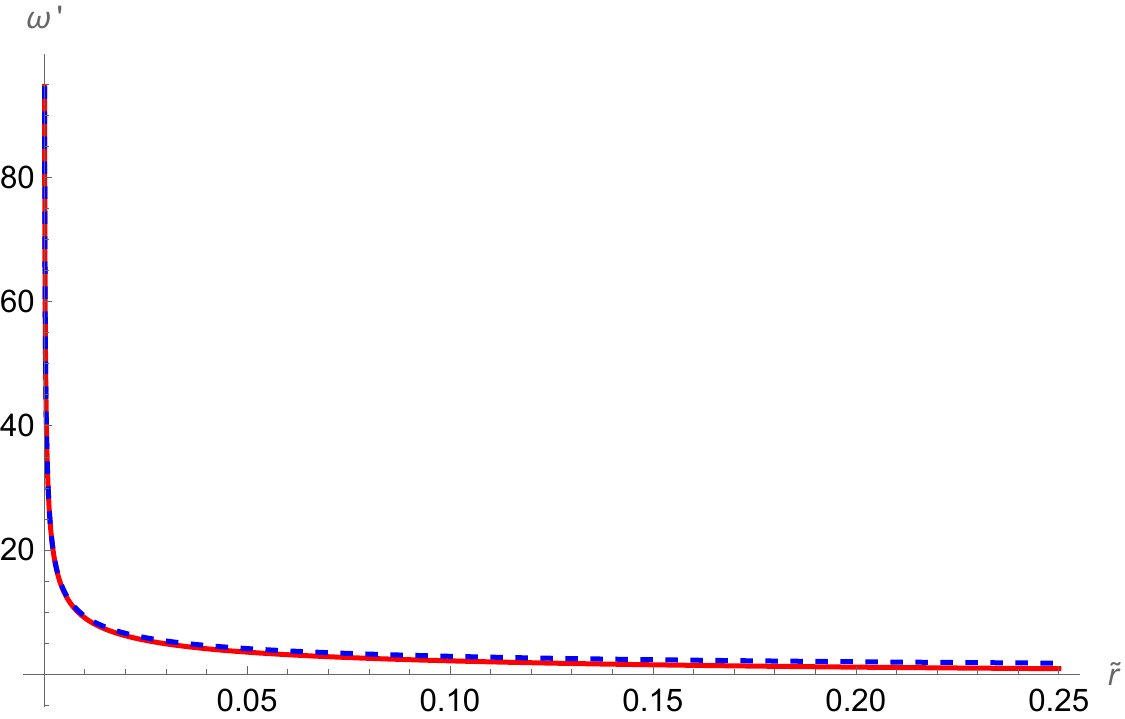}
		\vspace{4ex}
	\end{subfigure} 
	\begin{subfigure}[b]{0.5\linewidth}
		\centering
		\includegraphics[width=0.75\linewidth]{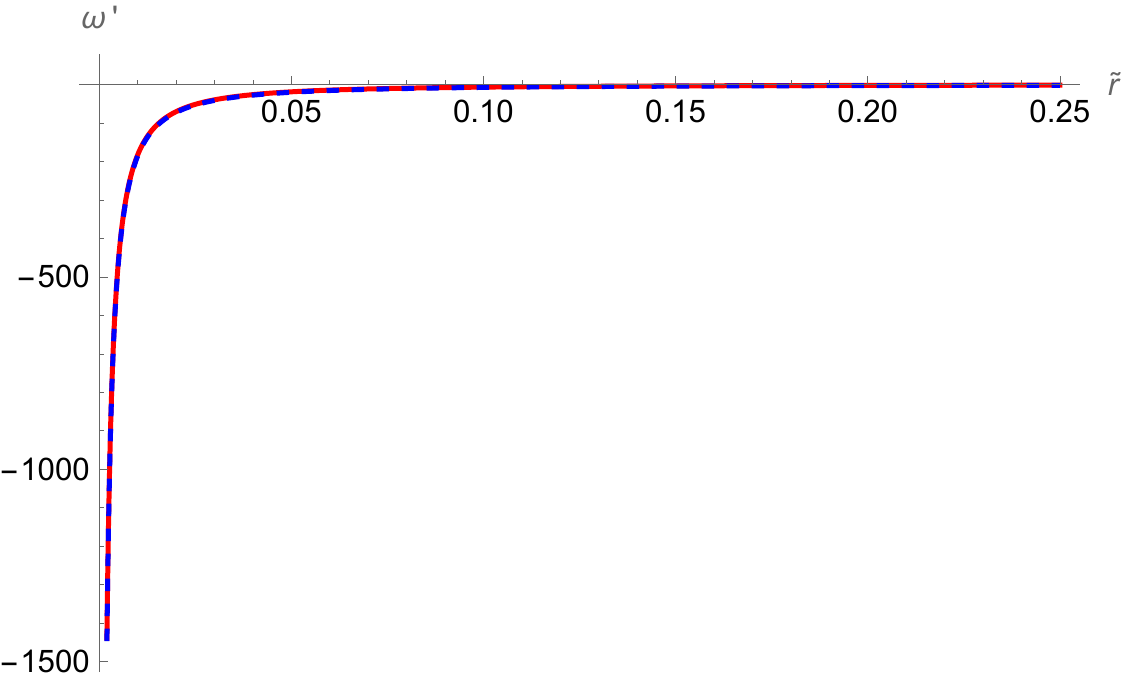} 
	\end{subfigure}
	\begin{subfigure}[b]{0.5\linewidth}
		\centering
		\includegraphics[width=0.75\linewidth]{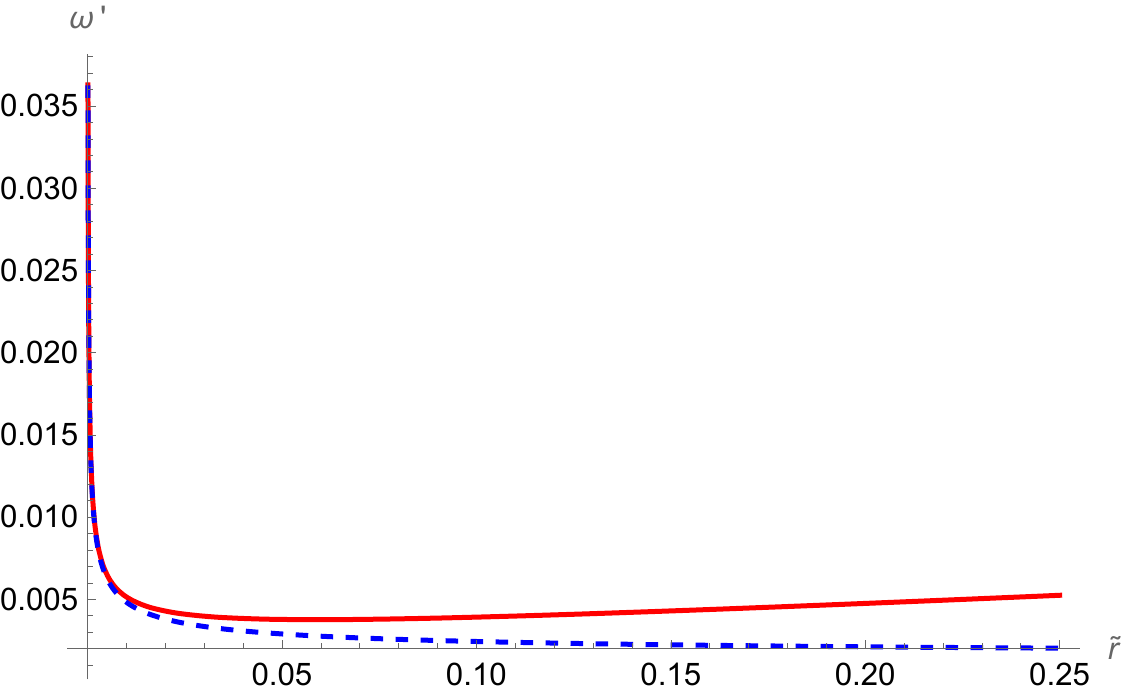} 
	\end{subfigure} 
 \caption{Four distinct numerical solutions for $\omega^{\prime}$ with parameters $q_m = 3$ and $\tilde{\mathcal{E}} = 0.05$. The blue dotted lines represent the asymptotic analytical behavior of $\omega'$. Here, $\omega$ is given by Eq.~\eqref{eq: SU2 omega for small tr}.}
\label{fig: mathc at zero der}
\end{figure}

Considering Taylor expansion for small-$\tilde{r}$ in Eq.~\eqref{eq: for w}, one can show that the asymptotic solution generally takes the following form
\begin{align}
     \label{eq: SU2 omega for small tr}
    \omega\big|_{\tilde{r} \to 0} = c_0 \pm \frac{2 (c_0^2 - 1) q_{m}}{(1 + q_{m}^2)^{1/2}} \sqrt{\tilde{r}} 
    + \frac{4 q_{m}^2  c_{0}{} \bigl( c_{0}^2 - 1\bigr)}{1 + q_{m}^2} \tilde{r}\;,
\end{align}
where $c_0$ is an arbitrary constant.
It is interesting to note that for small $\tilde{r}$ the solution is regular, while its derivative diverges. As shown in Figs.~\ref{fig: mathc at zero} and \ref{fig: mathc at zero der}, the above asymptotic solution and its first derivative perfectly match the numerical results.

However, for one of the four numerical solutions, the asymptotic expression~\eqref{eq: SU2 omega for small tr} fails to reproduce the observed behavior. Nevertheless, the analytical asymptotic behavior can still be obtained by considering a simplified form of Eq.~\eqref{eq: for w}. Specifically, for small values of $\tilde{r}$ and $\omega$, Eq.~\eqref{eq: for w} can be approximated as  
\begin{align*}
(1 + q_{m}^{2})\, \tilde{r}\, \omega_{\tilde{r}}^{2} - q_{m}^{2}\, \omega^{4} = 0 \;,
\end{align*}
which admits the following solutions:
\begin{align}\label{eq: asymp for div sol} 
\frac{\sqrt{1 + q_{m}^{2}}}{
c_{4}\, \sqrt{1 + q_{m}^{2}} \pm 2 q_{m} \sqrt{\tilde{r}}
} \;,
\end{align}
where $c_{4}$ is an arbitrary constant. By matching this asymptotic solution to the corresponding numerical result, the value of $c_{4}$ can be determined, thereby providing an accurate description of the observed numerical behavior.

\newpage

\bibliographystyle{JHEPmod}
\bibliography{refs}

\end{document}